\documentclass[prd]{revtex4}
\usepackage{epsf,latexsym}
\usepackage{amsmath,amssymb,amsthm}
\usepackage{graphicx}
\usepackage[all]{xy}
\usepackage{slashed}

\newcommand{\rxy}[1]{{\begin{xy}
0;<2mm,0mm>:<0mm,2mm>::0;0,#1\end{xy}}}

\newtheorem{thm}{Theorem}[section]

\newtheorem{defn}[thm]{Definition}

\newcommand{\bb}{\begin{eqnarray}}

\newcommand{\ee}{\end{eqnarray}}

\newcommand{\eee}{\nonumber\end{eqnarray}}

\newcommand{\pp}[1]{\begin{pmatrix} #1 \end{pmatrix}}

\newcommand{\beq}{\begin{equation}}

\newcommand{\eeq}{\end{equation}}

\newcommand{\bea}{\begin{eqnarray}}

\newcommand{\eea}{\end{eqnarray}}

\newcommand{\T}{{\rm tr}}

\newcommand{\ddfm}{\hbox{$^{\hat{f}}$\hspace{-0.15cm}
$\mathcal{D}$}}

\newcommand{\ddf}{\hbox{$^f$\hspace{-0.15cm} $\mathcal{D}$}}

\def\sv{\left<\sigma v\right>}

\def\eV{\,{\rm eV}}

\def\keV{\,{\rm keV}}

\def\MeV{\,{\rm MeV}}

\def\GeV{\,{\rm GeV}}

\def\s{{\,\rm s}}

\def\cm{{\,\rm cm}}

\def\g{{\,\rm g}}

\newcommand{\rxyz}[2]{{\begin{xy} 0;<2mm,0mm>:<0mm,2mm>::0;0,
,(5,-2)*{a}
,(10,-1.8)*{\bar{a}}
,(15,-2)*{b}
,(20,-2)*{c}
,(25,-2)*{d}
,(30,-2)*{e}
,(35,-1.8)*{\bar{e}}
,(40,-2)*{f}
,(2,-5)*{a}
,(2,-10)*{\bar{a}}
,(2,-15)*{b}
,(2,-20)*{c}
,(2,-25)*{d}
,(2,-30)*{e}
,(2,-35)*{\bar{e}}
,(2,-40)*{f}
,(5,-5)*\cir(#1,0){}
,(10,-5)*\cir(#1,0){}
,(15,-5)*\cir(#1,0){}
,(20,-5)*\cir(#1,0){}
,(25,-5)*\cir(#1,0){}
,(30,-5)*\cir(#1,0){}
,(35,-5)*\cir(#1,0){}
,(40,-5)*\cir(#1,0){}
,(5,-10)*\cir(#1,0){}
,(10,-10)*\cir(#1,0){}
,(15,-10)*\cir(#1,0){}
,(20,-10)*\cir(#1,0){}
,(25,-10)*\cir(#1,0){}
,(30,-10)*\cir(#1,0){}
,(35,-10)*\cir(#1,0){}
,(40,-10)*\cir(#1,0){}
,(5,-15)*\cir(#1,0){}
,(10,-15)*\cir(#1,0){}
,(15,-15)*\cir(#1,0){}
,(20,-15)*\cir(#1,0){}
,(25,-15)*\cir(#1,0){}
,(30,-15)*\cir(#1,0){}
,(35,-15)*\cir(#1,0){}
,(40,-15)*\cir(#1,0){}
,(5,-20)*\cir(#1,0){}
,(10,-20)*\cir(#1,0){}
,(15,-20)*\cir(#1,0){}
,(20,-20)*\cir(#1,0){}
,(25,-20)*\cir(#1,0){}
,(30,-20)*\cir(#1,0){}
,(35,-20)*\cir(#1,0){}
,(40,-20)*\cir(#1,0){}
,(5,-25)*\cir(#1,0){}
,(10,-25)*\cir(#1,0){}
,(15,-25)*\cir(#1,0){}
,(20,-25)*\cir(#1,0){}
,(25,-25)*\cir(#1,0){}
,(30,-25)*\cir(#1,0){}
,(35,-25)*\cir(#1,0){}
,(40,-25)*\cir(#1,0){}
,(5,-30)*\cir(#1,0){}
,(10,-30)*\cir(#1,0){}
,(15,-30)*\cir(#1,0){}
,(20,-30)*\cir(#1,0){}
,(25,-30)*\cir(#1,0){}
,(30,-30)*\cir(#1,0){}
,(35,-30)*\cir(#1,0){}
,(40,-30)*\cir(#1,0){}
,(5,-35)*\cir(#1,0){}
,(10,-35)*\cir(#1,0){}
,(15,-35)*\cir(#1,0){}
,(20,-35)*\cir(#1,0){}
,(25,-35)*\cir(#1,0){}
,(30,-35)*\cir(#1,0){}
,(35,-35)*\cir(#1,0){}
,(40,-35)*\cir(#1,0){}
,(5,-40)*\cir(#1,0){}
,(10,-40)*\cir(#1,0){}
,(15,-40)*\cir(#1,0){}
,(20,-40)*\cir(#1,0){}
,(25,-40)*\cir(#1,0){}
,(30,-40)*\cir(#1,0){}
,(35,-40)*\cir(#1,0){}
,(40,-40)*\cir(#1,0){}
#2\end{xy}}}

\newcommand{\rxya}[2]{{\begin{xy} 0;<2mm,0mm>:<0mm,2mm>::0;0,
,(5,-2)*{a} ,(10,-2)*{b} ,(2,-5)*{a} ,(2,-10)*{b} ,(5,-5)*\cir(#1,0){}
,(10,-5)*\cir(#1,0){} ,(5,-10)*\cir(#1,0){} ,(10,-10)*\cir(#1,0){}
#2\end{xy}}}

\newcommand{\rxycc}[2]{{\begin{xy} 0;<2mm,0mm>:<0mm,2mm>::0;0,
,(5,-2)*{\mu_{ij}} ,(10,-2)*{\mu_{ik}} ,(2,-5)*{\mu_{ij}}
,(2,-10)*{\mu_{\ell j}} ,(5,-5)*\cir(#1,0){} ,(10,-5)*\cir(#1,0){}
,(5,-10)*\cir(#1,0){} ,(10,-10)*\cir(#1,0){} #2\end{xy}}}

\begin{document}

\title{Composite Dark Matter with Invisible Light from Almost-Commutative Geometry}

\author{M.Yu.~Khlopov}
\email{Maxim.Khlopov@roma1.infn.it}
\address{Centre for CosmoParticle Physics "Cosmion", 125047,
Moscow, Russia\\
Moscow Engineering Physics Institute, 115409 Moscow, Russia
}

\author{C.A.~Stephan}
\email{christoph.stephan@cpt.univ-mrs.fr}
\address{Centre de Physique Theorique, CNRS-Luminy Case 907, \\
13288 Marseille cedex 9, France}
\email{christoph.stephan@cpt.univ-mrs.fr}

\begin{abstract}
Almost commutative geometry offers a specific way to unify general
relativity, quantum mechanics and gauge symmetries. The AC-model of
elementary particles, arising on this way, naturally embeds the
Standard model and predicts doubly charged AC-leptons, anion-like
$A^{--}$ and cathion-like $C^{++}$, which can bind in WIMP-like
(AC)-atoms, being a nontrivial candidate for cosmological dark
matter. This state is reached in the early Universe along a tail
of more manifest secondary frozen blocks. They should be now here
polluting the surrounding matter. The main secondary relics are
$C^{++}$ "anomalous helium" and a bound system of $A^{--}$ with an
ordinary helium ion $(^4He)^{++}$, which is able to attract and
capture (in the \emph{first three minutes}) all the free $A^{--}$
fixing them into a neutral \emph{OLe-helium} $(OHe)$ nuclear
interacting "atom" $(^4He^{++}A^{--})$. The model naturally
involves a new $U(1)$ gauge interaction, possessed only by the AC-leptons
and providing a Coulomb-like attraction between them. This
attraction stimulates the effective $A-C$ recombination into AC-atoms
inside dense matter bodies (stars and planets), resulting in a
decrease of anomalous isotopes below the experimental upper
limits. \emph{OLe-helium}
 pollution of terrestrial matter and
$(OHe)$ catalysis of nuclear reactions in it is one of the
exciting problems (or advantages?) of the present model.
\end{abstract}

\maketitle

\newpage
The problem of the  of new leptons, being among the most
important in  modern high energy physics, has acquired recently an
interesting cosmological aspect. New heavy leptons may be
sufficiently long-living to represent a new stable form of matter
and even to offer a nontrivial solution for cosmological dark
matter problem. At the present there are at least three main
elementary particle frames for Heavy Stable Quarks and Leptons and
their cosmological impact: (a) A fourth generation with heavy
a stable U-Quark and a neutral Lepton (neutrino) (above half the
Z-Boson mass)\cite{Shibaev}, \cite{Sakhenhance}, \cite{Fargion99},
\cite{Grossi}, \cite{Belotsky}, \cite{BKS} ; see also
 \cite{Berezhiani:1995du,Okun,Volovik:2003kh},
\cite{4had}, offering a possibility of an atom-like bound state
$[^4He^{++}(\bar U \bar U \bar U)^{--}]$ as a specific
nuclear-interacting candidate for dark matter \cite{anutium}; (b)
A Glashow's "Sinister" tera-U-quark and a tera-electron, whose
atomic bound states may be the dominant dark matter
\cite{Glashow,Fargion:2005xz} and (c) the possibility of doubly
charged AC-leptons\footnote{These particles were called $E^{--}$
and $P^{++}$ in \cite{leptons}. To avoid a misleading analogy with
the single charged electron and the proton we refer to them as to
Anion-Like EXotic Ion of Unknown Matter (ALEXIUM) $A^{--}$ and to
Cathion-Like EXotic Ion of Unknown Matter (CoLEXIUM) $C^{++}$.}
$A^{--}$ and $C^{++}$ recently revealed in \cite{leptons} for
the AC-model \cite{5}, following from the approach of
almost-commutative geometry by Alain Connes \cite{book}.

This latter option of the AC-model provides dark matter in the
form of evanescent bound AC-leptonic "atoms" $(AC)$ and can avoid
most of troubles of atom-like composite dark matter scenarios, if
the AC-leptons possess an additional new $U(1)$ gauge interaction
\cite{leptons}.
We shall address here our attention to the physical grounds of
the AC-model, on the conditions under which it gives rise to new
$U(1)$ interaction for the AC-leptons and on the self-consistency of
the cosmological dark matter scenario, based on this approach.

The AC-model \cite{5} appeared as a realistic elementary particle
model, based on the specific approach of \cite{book} to unify
general relativity and gauge symmetries. This
realization naturally embeds the Standard Model and extends its
fermion content by two heavy particles with opposite
electromagnetic and Z-boson charges. Having no other gauge charges
of the Standard model, these particles (AC-fermions) behave as heavy
stable leptons with charges $-2e$ and $+2e$, called here $A$ and
$C$, respectively. The mass of the AC-fermions has a "geometrical"
origin and is not related to the Higgs mechanism.
Here "geometrical" does not refer to the Planck mass, but is
attributed to the fact, that massive particles constitute an essential
contribution to the whole geometric framework of noncommutative
geometry.
In the absence
of AC-fermion mixing with light fermions, the AC-fermions can be
absolutely stable. Such absolute stability naturally follows from
strict conservation of the additional $U(1)$ gauge charge, which we
call $y$-charge, ascribed to AC-leptons, and we explore this
possibility in the present paper.

If the AC-leptons $A$ and $C$ have equal and opposite sign
of $y$-charges, strict conservation of the $y$-charge does not prevent
the generation of $A$ and $C$ excess; the excess of $A$ being equal to
the excess of $C$, as required in the further cosmological treatment.

AC-fermions are sterile relative to $SU(2)$ electro-weak
interaction, and do not contribute to the standard model
parameters.

Being absolutely stable, primordial heavy AC-leptons
should be present in modern matter \footnote{The mechanisms of
production of (meta)stable $Q$ (and $\bar Q$) hadrons and
tera-particles in the early Universe, cosmic rays and accelerators
were analyzed in \cite{4had,Fargion:2005xz,anutium} and the
possible signatures of their wide variety and existence were
revealed.}.

In the model \cite{5} the properties of heavy AC-fermions are
fixed by the almost-commutative geometry and the physical
postulates given in \cite{1}. The freedom resides in the choice of
the hyper-charge and the masses. According to this model
negatively charged $A^{--}$ and positively charged $C^{++}$ are
stable and
may form a neutral most probable and stable (while being
evanescent) $(AC)$ "atom". The AC-gas of such "atoms" is an ideal
candidate \cite{5,leptons} for a very new and fascinating dark
matter (like it was tera-helium gas in
\cite{Glashow,Fargion:2005xz}); because of their peculiar
WIMP-like interaction with matter they may also rule the  stages
of gravitational clustering in early matter dominated epochs,
creating first gravity seeds for galaxy formation.

However, in analogy to D, $^3$He and Li relics that are the
intermediate catalyzers of $^4$He formation in the Standard Big Bang
Nucleosynthesis (SBBN) and are important cosmological tracers of
this process, the AC-lepton relics from intermediate stages of a
multi-step process towards a final $(AC)$ formation must survive
with high abundance of {\it visible} relics in the present
Universe. We enlisted, revealed and classified such tracers, their
birth place and history up to now in  \cite{leptons}.

We found in  \cite{leptons} that $(eeC^{++})$
 should be here to remain among
us and its abundance should be strongly reduced in terrestrial
matter to satisfy known severe bounds on anomalous helium. This
reduction is catalyzed by relic neutral OLe-helium (named so from
\emph{O-Le}pton- \emph{helium}) $(^4He^{++}A^{--})$, because the
primordial component of free anion-like AC-leptons $A^{--}$ are
mostly trapped in the first three minutes into this puzzling
\emph{OLe-helium} "atom" $(^4He^{++}A^{--})$ with nuclear
interaction cross section, which provides anywhere an eventual later
$(AC)$ binding. This surprising catalyzer with screened Coulomb
barrier can influence the chemical evolution of ordinary matter,
but it turns out that the dominant process of OLe-helium
interaction with nuclei is quasi-elastic and might not result in copious
creation of anomalous isotopes. Inside dense matter objects (stars
or planets) its recombination with $(eeC^{++})$ into $(AC)$ atoms
can provide a mechanism for the formation of dense $(AC)$ objects.
We have mentioned in \cite{leptons} that most of the problems of
AC-cosmology, related with possible fractionating of $(eeC^{++})$
and OLe-helium owing to the strong difference in their mobility in
ordinary matter objects, can be avoided, if $A$ and $C$ possess
an additional U(1) gauge charge (y-charge). The Coulomb-like attraction
of y-charges prevents fractionating of anomalous helium and
OLe-helium and makes them recombine effectively into $(AC)$
atoms.

In the present paper we give some ideas on the exciting flavor of
the unification based on almost commutative geometry and on the
way it fixes the choice for the gauge symmetry group, underlying the
AC-model \cite{5,leptons} (Section \ref{flavor}) and the properties of
the AC-leptons $A$ and $C$, predicted by it (Section \ref{model}). We
consider their evolution in the early Universe and notice (Section
\ref{primordial}) that in spite of the assumed excess of particles
($A^{--}$ and $C^{++}$) the abundance of frozen out antiparticles
($\bar A^{++}$ and $\bar C^{--}$) is not negligible, as well as
a significant fraction of $A^{--}$ and $C^{++}$ remains unbound,
when $AC$ recombination takes place and most of AC-leptons form
$(AC)$ atoms. This problem of an unavoidable over-abundance of
by-products of "incomplete combustion" is unresolvable for models,
assuming dark matter, composed of atoms, binding single charged
particles, as it was revealed in \cite{Fargion:2005xz} for the sinister
Universe \cite{Glashow}. As soon as $^4He$ is formed in the Big Bang
nucleosynthesis it captures all the free negatively charged heavy
particles (Section \ref{primordial}). If the charge of such
particles is -1e (as it was the case for tera-electron in
\cite{Glashow}) positively charged ion $(^4He^{++}E^{-})^+$ puts
up a Coulomb barrier for any successive decrease of abundance of
the species, over-polluting by anomalous isotopes the modern Universe.
The double negative charge of $A^{--}$ in the considered AC-model
\cite{leptons} provides the binding with $^4He^{++}$ into a neutral
Ole-helium state, which catalyzes in the first three minutes an effective
binding into $(AC)$ atoms and a complete annihilation of the
antiparticles. Products of annihilation do not cause undesirable effects,
neither in the CMB spectrum, nor in light element abundances. Due to
early decoupling from the relativistic plasma y-photon background is
suppressed and its contribution to the total density in the period
of Big Bang Nucleosynthesis is compatible with observational
constraints.

Still, though the CDM in the form of $(AC)$ atoms is successfully
formed, $A^{--}$ (bound in OLe-helium) and $C^{++}$ (forming
anomalous helium atom $(eeC^{++})$) should be also present in the
modern Universe and the abundance of primordial $(eeC^{++})$ is by
up to {\it ten} orders of magnitude higher, than experimental
upper limits on the anomalous helium abundance in terrestrial
matter. This problem can be solved by OLe-helium catalyzed $(AC)$
binding of $(eeC^{++})$ (Subsection \ref{Matter}), but different
mobilities in matter of atomic interacting $(eeC^{++})$ and
nuclear interacting $(OHe)$ lead to the fractionating of these
species, preventing an effective decrease of the anomalous helium
abundance. We show that the $U(1)$ charge neutrality condition
naturally prevents this fractionating, making $(AC)$ binding
sufficiently effective to suppress a terrestrial anomalous isotope
abundance below the experimental upper limits.

However, though the $(AC)$ binding is not accompanied by strong
annihilation effects, as it was the case for 4th generation
hadrons \cite{4had}, gamma radiation from it inside large volume
detectors should take place. We clarify the astrophysical
uncertainties in estimation of the expected effect.

In this way AC-cosmology escapes most of the troubles, revealed
for other cosmological scenarios with stable heavy charged
particles \cite{4had,Fargion:2005xz} and provides a realistic
scenario for composite dark matter in the form of evanescent
atoms, composed by heavy stable electrically charged particles,
bearing the source of invisible light.

We give technical details for the approach to the particle theory,
based on almost-commutative geometry, in Appendices 1-5.

\section{\label{flavor} A flavor of almost-commutative geometry}

In the last few years several approaches to include the idea of
noncommutative spaces into physics have been established. One of
the most promising and mathematically elaborated is Alain Connes
{\it noncommutative geometry} \cite{book} where the main idea is
to translate the usual notions of manifolds and differential
calculus into an algebraic language. Here we will mainly focus on
the motivations why noncommutative geometry is a novel point of
view of space-time, worthwhile to be considered by theoretical
physics. We will furthermore try to give a glimpse on the main
mathematical notions (for computational details see appendices 1
to 4), but refer to \cite{book} and \cite{costa} for a thorough
mathematical treatment and to \cite{cc} and \cite{schuck} for its
application to the standard model of particle physics.

Noncommutative geometry has its roots in quantum mechanics and
goes back to Heisenberg \footnote{According to Roman Jackiw
\cite{Jackiw} Julius Wess told and documented the following: {\it
Like many interesting quantal ideas, the notion that spatial
coordinates may not commute can be traced to Heisenberg who, in a
letter to Peierls, suggested that a coordinate uncertainty
principle may ameliorate the problem of infinite self-energies.
... Evidently, Peierls also described it to Pauli, who told it to
Oppenheimer, who told it to Snyder, who wrote the first paper on
the subject \cite{Snyder}}} or even Riemann \cite{Riemann}. In the
spirit of quantum mechanics it seems natural that space-time
itself should be equipped with an uncertainty. The coordinate
functions of space-time should be replaced by a suitable set of
operators, acting on some Hilbert space with the dynamics defined
by a Dirac operator. The choice of a relativistic operator is
clear since the theory ought to be Lorentz invariant. As for the
Dirac operator, in favor of the Klein-Gordon operator, matter is
built from Fermions and so the Dirac operator is privileged. This
approach, now known as noncommutative geometry, has been worked
out by Alain Connes \cite{book}. He started out on this field to
find a generalized understanding to cope with mathematical objects
that seemed geometrical, yet escaped the standard approaches. His
work has its predecessors in Gelfand and Naimark \cite{Gelfand},
who stated that the topology of a manifold is encoded in the
algebra of complex valued functions over the manifold. Connes
extended this theorem and translated the whole set of geometric
data into an algebraic language. The points of the manifold are
replaced by the pure states of an algebra, which, inspired by quantum mechanics, acts
on a Hilbert space. With help of a Dirac operator acting as well
on the Hilbert space, Connes formulated a set of axioms which
allows to recover the geometrical data of the manifold. These
three items, the algebra, the Hilbert space and the Dirac operator
are called a spectral triple. But it should be noted that the set
of manifolds, i.e. space-times, which allow to be described by a
spectral triple is limited. These manifolds have to be Riemannian,
i.e. of Euclidean signature, and they have to admit a spin structure,
which is not true for any manifold. The second condition
presents no drawback since
space-time  falls exactly into this class. But asking the manifold
to be Euclidean, whereas special relativity requires a Lorentzian
signature, poses a problem, which is still open. Nevertheless one
can argue, along the line of Euclidean quantum field theory  that
this can be cured by Wick rotations afterwards.

A strong point in favor of the spectral triple approach is,
as the name noncommutative geometry already implies, that the
whole formulation is independent of the commutativity of the
algebra. So even when the algebra is noncommutative it is possible
to define a geometry in a consistent way. But then the geometry
gets equipped with an uncertainty relation, just as in quantum
mechanics.  With this generalization comes a greater freedom to
unify the two basic symmetries of nature, namely the
diffeomorphism invariance (= invariance under general coordinate
transformations) of general relativity and the local gauge
invariance of the standard model of particle physics. In the case
of ordinary manifolds theorems by O'Raifeartaigh \cite{ORaif},
Coleman and Mandula \cite{Coleman} as well as the theorem of
Mather \cite{Mather} prohibit such a unification (for details see
\cite{book}).

The standard model can be constructed as a classical  gauge theory
which describes the known elementary particles by means of the
symmetries they obey, together with the electro-weak and the
strong force. In contrast to general relativity, this classical
field theory allows to pass over to a quantum field theory. All
elementary particles are fermions and the forces acting between
them are mediated by bosons. The symmetries of the theory are
compact Lie groups, for the standard model of particle physics the
underlying symmetry goup
is $U(1)\times SU(2) \times SU(3)$. Fermions are Dirac spinors,
placed in multiplets which are representations of the symmetry
groups. A peculiar feature of the standard model is that fermions
are chiral. This poses a serious problem, since mass terms mixing
left- and right-handed states would explicitly break the symmetry.
To circumvent this an extra boson, the Higgs boson, has to be
introduced. In the widely used formulation of the standard model
this Higgs mechanism has to be introduced by hand. All the
non-gravitational forces and all known matter is described in a
unified way. But it is not possible to unify it on
the footing of differential geometry with general relativity. The
problem is that no manifold exists, which has general coordinate
transformations and a compact Lie group as its diffeomorphism
group. But here the power of noncommutative geometry comes in.

The first observation is that the general coordinate
transformations of a manifold correspond to the automorphisms
\footnote{An algebra automorphism is a bijective map from the
algebra into itself, which preserves the whole structure of the
algebra (i.e. the addition, the multiplication and, if present,
the involution). The algebra automorphisms form a group.} of the
algebra of complex valued functions over the manifold. Chamseddine
and Connes \cite{cc} discovered that it is possible to define an
action, called the {\it spectral action}, to give space-time in
the setting of spectral triples a dynamics, just as the
Einstein-Hilbert action for general relativity. This spectral
action is given by the number of eigenvalues of the Dirac operator
up to a cut-off. It is most remarkable that this action reproduces
the Einstein-Hilbert action in the limit of high eigenvalues of
the Dirac operator. The crucial observation is now that in
contrast to the diffeomorphisms of a manifold, the automorphisms
of an algebra allow to be extended to include compact Lie groups.
These are the automorphisms of matrix algebras. And since the
whole notion of a spectral triple is independent of the
commutativity of the algebra, it is possible to combine the
algebra of functions over the space-time manifold with an algebra
being the sum of simple matrix algebras by tensorising. These
combined function-matrix geometries are called {\it
almost-commutative} geometries. The part of the spectral triple
based on the matrix algebra is often called the {\it finite} or
{\it internal part}. Indeed, they contain an infinite number of
commutative degrees of freedom plus a finite number of
noncommutative ones. The former are outer, the latter are
inner automorphisms.

To see how the Higgs scalar, gauge potentials and gravity  emerge
one starts out with an almost-commutative spectral triple over a
flat manifold $M$. The corresponding algebra of complex valued
functions over the manifold will be $\mathcal{A}_R$ (where the
subscript $R$ stands for {\it Riemannian} \footnote{These
subscripts will be dropped when no confusion will arise from
it.}), the Hilbert space $\mathcal{H}_R$ is the Hilbert space of
Dirac spinors and the Dirac operator is simply the flat Dirac
operator $\slashed{\partial}$. As mentioned above, the
automorphisms of the algebra $\mathcal{A}_R$ coincide with the
diffeomorphisms, i.e. the general coordinate transformations, of
the underlying manifold, Aut$(\mathcal{A}_R)=$Diff$(M)$. To render
this function algebra noncommutative, a matrix algebra
$\mathcal{A}_f$ (where the subscript $f$ stands for {\it finite})
is chosen. The exact form of this matrix algebra is of no
importance for the moment (as long as its size is at least
two). The Hilbert space $\mathcal{H}_f$ is finite dimensional and
the Dirac operator $\mathcal{D}_f$ is a complex valued matrix. For
the detailed form of the internal Dirac operator see Appendix 1.

It is a pleasant feature of spectral triples that the tensor
product of two spectral triples is again a spectral triple. So
building the tensor product one finds for the algebra and the
Hilbert space of the almost commutative geometry
\begin{equation}
\mathcal{A}_{AC} = \mathcal{A}_R \otimes \mathcal{A}_f, \quad
\mathcal{H}_{AC} = \mathcal{H}_R \otimes \mathcal{H}_f.
\end{equation}
The Dirac operator needs a little bit more care to comply with the
axioms for spectral triples. It is given by
\begin{equation}
\mathcal{D}_{AC} = \slashed{\partial} \otimes 1_f + \gamma^5
\otimes \mathcal{D}_f,
\end{equation}
where $1_f$ is a unity matrix whose size is  the size of the
finite Dirac operator $\mathcal{D}_f$ and $\gamma^5$ is
constructed in the standard way from the Dirac gamma matrices. The
automorphism group of the almost-commutative algebra
$\mathcal{A}_{AC}$ is the semi direct product of the
diffeomorphisms of the underlying manifolds and the gauged
automorphisms of the matrix algebra. For example, with the matrix
algebra $\mathcal{A}_f=M_2 (\mathbb{C})$ one would have the gauged
unitary group $SU(2)$ in the automorphism group. This is exactly
the desired form for a symmetry group.

Now these automorphisms
\begin{equation}
Aut(\mathcal{A}_{AC}) = Aut(\mathcal{A}_R) \ltimes
Aut(\mathcal{A}_f) \ni (\sigma_R, \sigma_f) \label{auto}
\end{equation}
have to be lifted (=represented) to the Hilbert space
$\mathcal{H}_{AC}$. This is necessary to let them act on the
Fermions as well as to fluctuate or gauge the Dirac operator. It
is achieved by the lift $L(\sigma_R, \sigma_f)$ which is defined
via the representation of the algebra on the Hilbert space. For
details see Appendix 4.

For the moment the Dirac operator $\mathcal{D}_{AC}$ consists of
the Dirac operator on a flat manifold and a complex valued matrix.
Now, to bring in the Higgs scalar, the gauge potentials and
gravity the Dirac operator has to be fluctuated or gauged with the
automorphisms (\ref{auto})
\bb
\ddf_{AC} &&:= L(\sigma_R, \sigma_f) \mathcal{D}_{AC}
L(\sigma_R, \sigma_f)^{-1}
\nonumber \\
&&= L(\sigma_R, \sigma_f) (\slashed{\partial} \otimes 1_f
)L(\sigma_R, \sigma_f)^{-1} +L(\sigma_R, \sigma_f) ( \gamma^5
\otimes \mathcal{D}_f)L(\sigma_R, \sigma_f)^{-1}
\nonumber \\
&&= \slashed{\partial}_{cov.} + \gamma^5 \otimes L(\sigma_f)
\mathcal{D}_f L(\sigma_f)^{-1}=  \slashed{\partial}_{cov.} + \gamma^5 \otimes
\ddf_{f}.
\ee
In the last step it turns out that
$\slashed{\partial}_{cov.}:=L(\sigma_R, \sigma_f)
(\slashed{\partial} \otimes 1_f )L(\sigma_R, \sigma_f)^{-1}$ is
indeed the covariant Dirac operator on a curved space time, when
the appearing gauge potentials have been promoted to arbitrary
functions, i.e. after applying Einstein's equivalence principle
(for details see \cite{cc}). $\slashed{\partial}_{cov.}$ has
automatically the correct representation of the gauge potentials
on the Hilbert space of Fermion multiplets. The gauge potentials
thus emerge from the usual Dirac operator acting on the gauged
automorphisms of the inner algebra.

As for the Higgs scalar, it is identified with $\ddf_{f}:=L(\sigma_f)
\mathcal{D}_f L(\sigma_f)^{-1}$. Here the commutative
automorphisms being the diffeomorphisms $\sigma_R$ of the manifold
drop out, since they commute with the matrix $  \mathcal{D}_f$.
This is not true for the gauged automorphisms $\sigma_f$ since
they are matrices themselves.

From the gauged Dirac operator $\ddf_{AC}$ the spectral action is calculated
via a heat-kernel expension to be the Einstein-Hilbert action plus the Yang-Mills-Higgs
action. The Higgs potential in its well known quartic form is a
result of this calculation. It should be pointed out that the heat-kernel
expansion is performed up to a cut-off and so the obtained Einstein-Hilbert action
and Yang-Mills-Higgs action should be considered as effective actions.
The details of the calculation of the
spectral action goes beyond the scope of this publication and we
refer again to \cite{cc} for a detailed account. For the internal
part $\ddf_{f}$ of the gauged Dirac operator $\ddf_{AC}$ the
spectral action gives exactly the Higgs potential
\bb
V(\ddf_{f} )= \lambda\  \T\!\left[ (\ddf_{f} )^4\right] -\textstyle{\frac{\mu
^2}{2}}\ \T\!\left[ (\ddf_{f}) ^2\right] ,
\ee
where $\lambda $ and $\mu $ are positive constants, as well as
the kinetic term for the Higgs potential.  To determine the a sensible
value for the cut-off in the heat-kernel expansion, it is instructive
to note, that at the cut-off the couplings of the non-abelian gauge groups
and the coupling $\lambda$ of the Higgs potential are closely tied together.

Choosing as matrix algebra $\mathcal{A}_f=\mathbb{C} \oplus
\mathbb{H} \oplus M_3(\mathbb{C})$, where $\mathbb{H}$ are the
quaternions, one recovers with a suitable choice for the Hilbert
space that the spectral action reproduces the Einstein-Hilbert
action and the Yang-Mills-Higgs action of the standard model.
The cut-off is then fixed to be at the energy where the coupling
$g_2$ of the weak group $SU(2)_L$ and the coupling $g_3$
of the colour group $SU(3)_C$ become equal ($\sim 10^{17}$GeV).
At the cut-off these two couplings and the Higgs coupling $\lambda$
are related as
\bb
g_3^2=g_2^2=3 \lambda.
\ee
Assuming a great dessert up to the cut-off,  this relation allows
to let the Higgs coupling run back to lower energies and
to calculate the Higgs mass a. A detailed calculation
can be found in \cite{schuck} and gives a Higgs mass
of $m_{Higgs}= 175.4 \pm 4.7$ GeV, where the uncertainty
is due to  the uncertainty in the top-quark mass.

Recapitulating, the
Higgs scalar together with its potential emerge naturally as the
"Einstein-Hilbert action" in the noncommutative part of the
algebra. Here it has become possible for the first time to give
the Higgs scalar a geometrical interpretation. In the
almost-commutative setting it plays at the same time the r\^ole of
the metric in the finite part of the geometry as well as that of
the fermionic mass matrix.

One may interpret an almost-commutative geometry as a kind of
Kaluza-Klein theory, where the extra dimensions are discrete.
These extra dimensions are produced by the matrix algebra and they
provide for extra degrees of freedom without being visible.
Furthermore the Yang-Mills-Higgs action can be viewed in the
almost-commutative setting as the gravitational action, or
Einstein-Hilbert analogue for the "discrete part" of space-time.
From this point of view the gauge bosons, i.e. the Higgs boson,
the Yang-Mills bosons and the graviton form a unified
"super-multiplet". But of course space-time in the classical sense
ceases to exist in noncommutative geometry, just as there is no
classical phase space in quantum mechanics. The space-time has
been replaced by operators and extended by discrete extra
dimensions.

\begin{figure}
\begin{center}
\includegraphics[width=11cm]{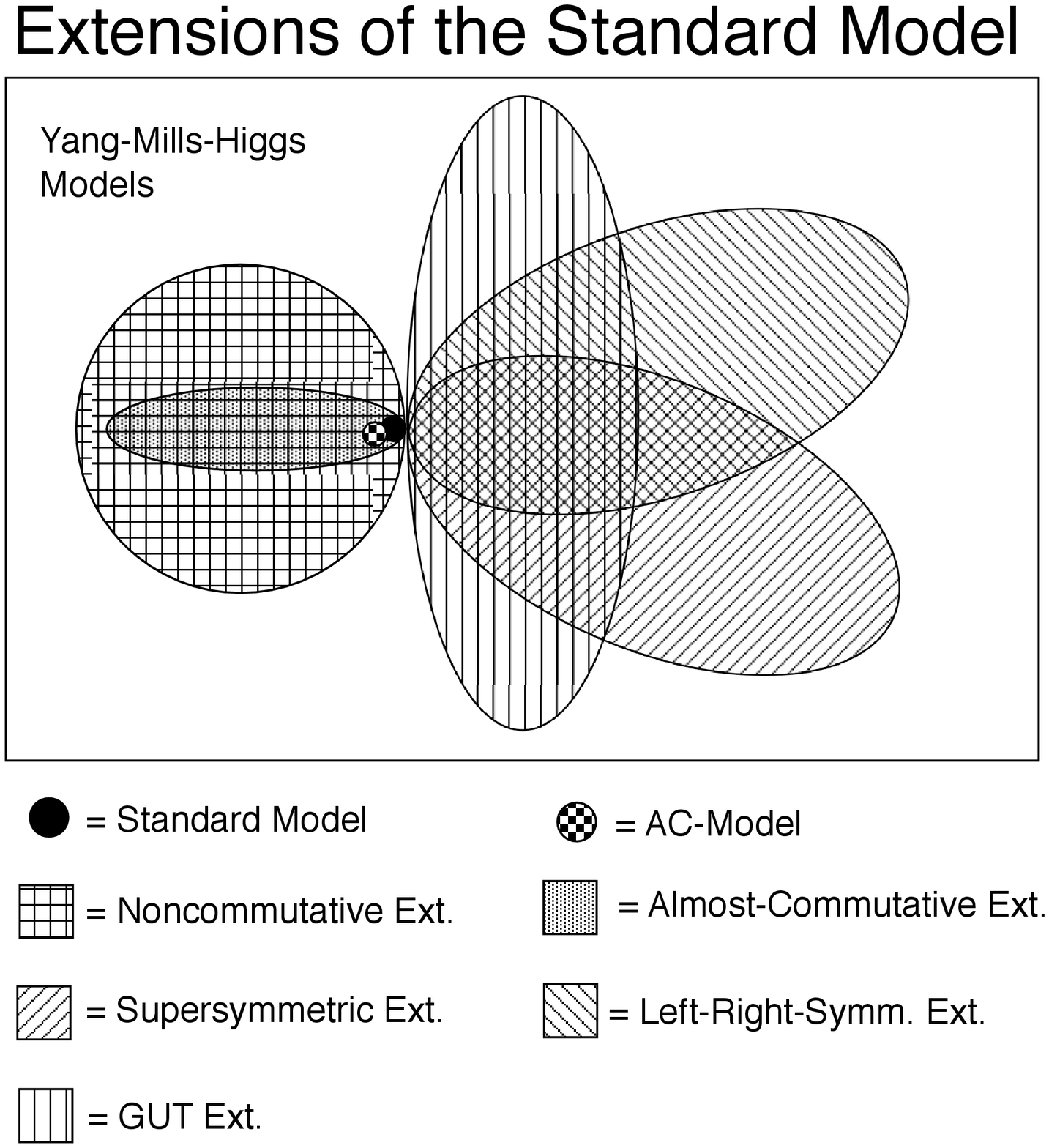}
\caption{Yang-Mills-Higgs models extending the Standard Model}
\label{versus}
\end{center}
\end{figure}

The immediate question that arises is: Which kind of
Yang-Mills-Higgs theory may fit into the frame work of
almost-commutative geometry? The set of all Yang-Mills-Higgs
theories is depicted in figure \ref{versus}. One sees that
left-right symmetric, grand unified and supersymmetric theories do
not belong to the elected group of noncommutative models.
But, as mentioned above, the standard model,
resulting from an almost-commutative geometry,
as well as the AC-model, do.

One of the main tasks of the present research in
almost-commutative geometry is to clarify the  structure of this
restricted sub-set of Yang-Mills-Higgs theories that originate
from  spectral triples. Since this is still an unscalable
challenge it is necessary to adopt a minimal approach. Imposing
certain constraints which are gathered from different areas
reaching from Riemannian geometry over high energy physics to
quantum field theory and starting out with only up to four
summands in the matrix algebra part of the almost commutative
geometry, one can give a classification from the particle
physicist's point of view. As it is custom in particle physics,
space-time curvature  will be neglected. Nonetheless the
Riemannian part of the spectral triple plays a crucial role in the
spectral action, introducing derivatives and thus the gauge
bosons. Setting the curvature to zero when the Einstein-Hilbert
and Yang-Mills-Higgs action have been obtained from the spectral
action leaves the Yang-Mills-Higgs action. With respect to this
part of the spectral action the classification will be done. As a
consequence only the finite matrix algebra part of the spectral
triple has to be classified since only the internal Dirac operator
enters into the Higgs scalar, as was shown above. The minimum of
the Higgs potential is the mass matrix of the fermions.

This classification proceeds in two steps. First all the possible
finite spectral triples, with a given number of summands of simple
matrix algebras, have to be found. This classification of finite
spectral triples has been done in the most general setting by
Paschke, Sitarz \cite{pasch} and Krajewski \cite{Kraj}. To
visualize a finite spectral triple Krajewski introduced a
diagrammatic notion, {\it Krajewski diagrams}, which
encode all the algebraic data of a spectral triple. For a more
detailed account see Appendix 3. If one imposes now as a first
condition that the spectral triple be irreducible, i.e. that the
finite Hilbert space be as small as possible, one is led to the
notion of a minimal Krajewski diagram. For a given number of
algebras, the algebra representation on the Hilbert space and the
possible Dirac operators are encoded in these diagrams by arrows
connecting two sub-representations. Finding the minimal diagrams
via this diagrammatic approach is very convenient and quite simple
for up to two summands in the matrix algebra. In this case only a
handful of diagrams exist and it is difficult to miss a diagram.
But with three and more algebras the task quickly becomes
intractable. For three algebras it may be done by hand, but one
risks to overlook some diagrams. It is thus fortunate that the
diagrammatic treatment allows to translate the algebraic problem
of finding spectral triples into the combinatorial problem of
finding minimal Krajewski diagrams. This can then be put into a
computer program. Still the problem is quite involved and the
algorithm to find minimal Krajewski diagrams needs a lot of care.
Furthermore the number of possible Krajewski diagrams increases
rapidly with the number of summands of matrix algebras and reaches
the maximal capacity of an up-to-date personal computer at  five
summands.

Nonetheless it is possible to find a Krajewski diagram with six
summands in the matrix algebra which is in concordance with the
physical requirements presented below. It is the aim of this paper
to evaluate its impact on the dark matter problem in cosmology.

If one has found the minimal Krajewski diagrams the second major
step follows.  From each Krajewski diagram all the possible
spectral triples have to be extracted. These are then analyzed
with respect to the following heteroclitic criteria:

\begin{itemize}
\item[$\bullet$] For simplicity and in view of the minimal
approach the spectral triple should be irreducible. This means
simply that the Hilbert space cannot be reduced while still
obeying all the axioms of a spectral triple. 
\item[$\bullet$]  The
spectral triple should be non-degenerate, which means that the
fermion masses should be non-degenerate, up to the inevitable
degeneracies which are left and right, particle and antiparticle
and a degeneracy due to a color. This condition has its origin in
perturbative quantum field theory and asserts that the possible
mass equalities are stable under renormalization flow.
\item[$\bullet$] Another criterion also stemming from quantum
field theory is that the Yang-Mills-Higgs models should be free of
Yang-Mills anomalies. In hope of a possible unified quantum theory
of all forces, including gravity, it is also demanded that the
models be free of mixed gravitational anomalies. 
\item[$\bullet$]
From particle phenomenology originates the condition that the
representation of the little group has to be complex in each
fermion multiplet, in order to distinguish particles from
antiparticles. 
\item[$\bullet$] The last item is the requirement
that massless fermions should be neutral under the little group.
This is of course motivated by the Lorentz force.
\end{itemize}

\noindent Now the Higgs potential has to be minimized and the
resulting models have to be compared with the above list of
criteria. If a model fits all the points of the list it may be
considered  of physical importance, otherwise it will be
discarded.

\section{\label{model} The particle model}
Among the possible almost-commutative Yang-Mills-Higgs models is
the standard model of particle physics with one generation of quarks
and leptons, for details we refer to
\cite{schuck}. It could furthermore be shown,  \cite{1,2,3,4},
that the standard model takes a most prominent position among
these Yang-Mills-Higgs models.

But the classification of almost-commutative geometries also
allows to go beyond the standard model in a coherent way. Here
heavy use is made of Krajewski diagrams \cite{Kraj}, which allow
to visualize the structure of almost-commutative geometries. The
particle model analyzed in the present publication is an extension
of the AC-lepton model presented in \cite{5} which was analyzed
with respect to its cosmological implications in \cite{leptons}.
It is remarkable to note that the almost-commutative geometry of
the basic AC-lepton model, which builds on the internal algebra
\bb \mathcal{A}= \mathbb{C} \oplus \mathbb{H} \oplus
M_3(\mathbb{C}) \oplus \mathbb{C} \oplus \mathbb{C}  \oplus
\mathbb{C}, \ee
allows to be enlarged to
\bb \mathcal{A}= \mathbb{C} \oplus M_2(\mathbb{C}) \oplus
M_3(\mathbb{C})  \oplus \mathbb{C} \oplus \mathbb{C}  \oplus
\mathbb{C} \ee
which produces through the so called centrally extended lift, for
details see \cite{farewell}, a second $U(1)$ gauge group in
addition to the standard model hypercharge group $U_Y(1)$. In
spirit with the previous nomenclature this group will be called
$U_{AC}(1)$, where AC stands again for
\underline{a}lmost-\underline{c}ommutative. Choosing the central
charges to reproduce the  standard model and the AC-particles with
the correct electric charges, a minimal extension consists in
coupling this new gauge group in the Lagrangian only to the
AC-fermions. For a detailed derivation from the corresponding
Krajewski diagram to the Lagrangian of the model we refer to
Appendix 5. The AC-particles do not participate in the Higgs
mechanism and consequently the AC-gauge group stays unbroken:
\bb U_Y(1)\times SU_L(2)\times SU_c(3)  \times U_{AC}(1)
\longrightarrow U_{em}(1)\times SU_c(3)  \times U_{AC}(1)
\nonumber \ee
The Lagrangian of the model consists of the usual standard model
Lagrangian, the  Lagrangian for the AC-particles and the new term
for the AC-gauge potential. We shall only give the two new parts
of the Lagrangian for the AC-fermion spinors $\psi_A$ and $\psi_C$
and the AC-gauge curvature $\tilde B_{\mu \nu}$:
\bb \mathcal{L}_{AC} &=& i \psi_{A L}^\ast D_A \psi_{A L} + i
\psi_{A R}^\ast D_A \psi_{A R} + m_A \psi_{A L}^\ast \psi_{A R} +
 m_A \psi_{A R}^\ast \psi_{A L}
\nonumber  \\
&&+ \; i \psi_{C L}^\ast D_C \psi_{C L} + i \psi_{C R}^\ast D_C
\psi_{C R} + m_C \psi_{C L}^\ast \psi_{C R} + m_C \psi_{C R}^\ast
\psi_{C L}
\nonumber \\
&& \; - \frac{1}{4} \tilde B_{\mu \nu} \tilde B^{\mu \nu}.
\nonumber \ee
The covariant derivatives $D_{A/C}$ and the gauge curvature are
given by
\bb D_{A/C} &=& \gamma^\mu \partial_\mu + \frac{i}{2} \, g' \,
Y_{A/C} \gamma^\mu B_\mu +  \frac{i}{2} \, g_{AC} \, \tilde
Y_{A/C}  \gamma^\mu \tilde B_\mu
\nonumber \\
&=&\gamma^\mu  \partial_\mu + \frac{i}{2} \, e \, Y_{A/C}
\gamma^\mu  A_\mu - \frac{i}{2} \, g' \, \sin \theta_w Y_{A/C}
\gamma^\mu  Z_\mu +  \frac{i}{2} \, g_{AC} \, \tilde Y_{A/C}
\gamma^\mu \tilde B_\mu, \nonumber \ee
and
\bb \tilde B_{\mu \nu} = \partial_\mu \tilde B_{\nu} -
\partial_\nu \tilde B_{\mu} \ee
As has been pointed out in \cite{leptons}, the electric charge
of the AC-leptons has to be $Q_{em} = \pm 2 e$, where $e$ is the
electric charge of the electron. Otherwise unwanted forms of
OLe-Helium ions would appear. This requires $Y_{A/C} = \mp 2$. For
simplicity the AC-hyper charge is also chosen to be $\tilde
Y_{A/C} = \mp 2$, but it cannot be fixed by almost-commutative
geometry. This  also applies to the AC-coupling $g_{AC}$ which
has to be fixed by experiment. If the coupling is chosen small
enough, the AC-fermions will exhibit a supplementary long range
force with a Coulomb like behavior. The corresponding necessarily massless
"photons"
will be called $y$-photons. Implications and effects on the high
energy physics of the standard model will not be considered in
this paper, but there may be detectable effects due to
interactions between AC-fermions and standard model particles on
loop level.

Indeed, loop diagrams with virtual $A$ and $C$ pairs induce mixing
between $y$-photon and ordinary gauge bosons ($y-\gamma$ and
$y-Z$). Due to this mixing ordinary particles acquire new long
range ($y$) interaction, which, however, can be masked in the
electro-neutral matter.

The masses of the standard model fermions are obtained by
minimizing the Higgs potential. It turns out that the masses $m_A$
and $m_C$ of the new fermions do not feel the fluctuations of the
Dirac operator and are thus Dirac masses which do not stem from
the Higgs mechanism but have a purely geometrical origin. This is
due to the necessarily vector like coupling of the gauge group induced by the
lift of the automorphisms. Consequently these Dirac masses do not
break gauge invariance. The mass scale will later be fixed on
cosmological grounds.

\section{\label{primordial} The cosmological model}

The model  \cite{5,leptons} admits that in the early Universe a
charge asymmetry of AC-fermions can be generated, as it is the
case for ordinary baryons, so that an $A$ and a $C$ excess saturates
the modern dark matter density, dominantly in the form of $(AC)$
atoms. For the baryon excess $\eta_b= n_{b\,mod}/n_{\gamma \,mod}
= 6 \cdot 10^{-10}$ it gives an AC-excess
 \beq
 \eta_{A}=n_{A\,mod}/n_{\gamma \,mod} = \eta_{C}=n_{C\,mod}/n_{\gamma \,mod} = 3 \cdot 10^{-11}
(\frac{100{\GeV}}{M}), \label{excess}
\eeq
where $M=m_A+m_C$ is
the sum of the masses of $A$ and $C$. Following
\cite{Glashow,Fargion:2005xz,leptons}, it is convenient to relate
the baryon $\Omega_b=0.044$ and the AC-lepton densities
$\Omega_{CDM}=0.224$ with the entropy density $s$ and to introduce
$r_b = n_b/s$ and $r_{A}=r_{C}=n_{A}/s=n_{C}/s$. Taking into
account that $s_{mod}=7.04\cdot n_{\gamma\,mod},$ one obtains $r_b
\sim 8 \cdot 10^{-11}$ and \beq r_{A} =r_{C} = 4 \cdot 10^{-12}
(\frac{100{\GeV}}{M}). \label{sexcess} \eeq We'll further assume
that $m_A=m_C=M/2=m$, so the AC -fermion excess Eq.(\ref{sexcess})
is given by \beq \kappa_{A} =\kappa_{C} =r_{A} -r_{\bar A} =r_{C}
-r_{\bar C}= 2 \cdot 10^{-12} (\frac{100{\GeV}}{m}) = 2 \cdot
10^{-12}/S_2, \label{Eexcess} \eeq where $S_2 = m/100{\GeV}$.

\subsection{\label{Chronology} Chronological cornerstones of the AC-Universe}
After the generation of AC-lepton asymmetry in chronological order
the thermal history of AC-matter follows the trend, which we have
thoroughly studied in \cite{leptons} for $m_A = m_C = m = 100
S_2{\GeV} $. Therefore we briefly outline here this trend,
specifying in more details the effects of the $y$-interaction

1) $10^{-10}S_2^{-2}\s \le t \le 6 \cdot10^{-8}S_2^{-2}\s$ at $m
\ge T \ge T_f=m/31 \approx 3 S_2 \GeV.$  AC-lepton pair $A \bar A$
and $C \bar C$ annihilation and freezing out take place. At $S_5
\le 5$ frozen out concentration of antiparticles is exponentially
suppressed, while the concentration of $A$ and $C$ tends to the
value of their excess (\ref{sexcess}).
For larger $m$ the abundance of frozen out AC-lepton pairs is not
suppressed in spite of AC-lepton excess and antiparticles survive
until later stages.

2) $1.5 \cdot 10^{-4}S_2^{-2}k_y^{-4}\s \le t \le 1.3 \cdot
10^{-1} S_2^{-2}k_y^{-4}\s$ at $I_{AC} \approx 80 S_2 k_y^{2}\MeV
\ge T \ge I_{AC}/30 .$ In this period recombination of negatively
charged AC-leptons $A^{--}$ with positively charged
$C^{++}$-leptons can lead to the formation of AC-lepton "atoms"
$(AC)$ with potential energy $I_{AC}= Z_A^2 Z_C^2\alpha^2 k_y^{2}m
\approx 80 S_2 k_y^{2}\MeV $ ($Z_A=Z_C=2$ and $k_y = (1 +
\alpha_y/(Z_AZ_C\alpha))/2 \approx 1$ for $\alpha_y \sim 1/30$).
Together with neutral $(AC)$-"atoms" free charged $A^{--}$ and
$C^{++}$ are also left, being the dominant form of AC-matter at
$S_2 > 6$.

3) $ t \sim 1.5 \cdot 10^{-4}S_2^{-2}k_y^{-4}\s$  at $T \sim I_{A}
= I_{C} = 80 S_2 k_y^2\MeV.$ The temperature corresponds to the
binding energy $I_{A} = I_{C} = Z_A^4 \alpha^2 k_y^2 m = Z_C^4
\alpha^2 k_y^2 m \approx 80 S_2 k_y^2 \MeV$ ($Z_A=Z_C=2$) of twin
AC-positronium "atoms" $(A^{--} \bar A^{++})$ and $(C^{++} \bar
C^{--})$, in which $\bar A^{++}$ and $\bar C^{--}$ annihilate. At
large $m$ this annihilation is not at all effective to reduce the
$A \bar A$ and $C \bar C$ pairs
abundance.
These pairs are eliminated in
the course of the successive evolution of AC-matter.

4) $100\s \le t \le 300\s$  at $100 \keV\ge T \ge I_{OHe}/27
\approx 60 \keV,$ where $I_{OHe}= Z_{He}^2 Z_{A}^2 \alpha^2
m_{He}/2 = 1.6 \MeV$ is the ionization potential of a
$(^4He^{++}A^{--})$ "atom". Helium $^4$He is formed in the
Standard Big Bang Nucleosynthesis and virtually all free $A^{--}$
are trapped by $^4$He in OLe-helium $(^4He^{++}A^{--})$.
Note that in the period $100 \keV \le T \le 1.6 \MeV$ helium
$^4$He is not formed, therefore it is only after {\it the first
three minutes}, when $(^4He^{++}A^{--})$ trapping of $A^{--}$ can
take place. Being formed, OLe-helium catalyzes the binding of free
$C^{++}$ with its constituent $A^{--}$ into $(AC)$-"atoms". In
this period free $\bar C^{--}$ are also captured by $^4$He. At
large $m$ effects of $(A^{--} \bar A^{++})$ and $(C^{++} \bar
C^{--})$ annihilation, catalyzed by OLe-helium, do not cause any
contradictions with observations.

The presence of new relativistic species - a gas of primordial
$y$-photons - does not influence the light element abundances, since
the $y$-photons decouple at $T< T_f (Z^2 \alpha/\alpha_y)$ from
the cosmological plasma after AC-lepton pairs are frozen out at
$T_f=m/30 \approx 3 S_2 \GeV.$ Here $\alpha_y$ is the fine
structure constant of the $y$-interaction and $Z=2$. Therefore
the contribution of $y$-photons into the total number of relativistic
species in the period of SBBN is suppressed.

\subsection{\label{EHESBBN} OLe-helium in the  SBBN}
OLe-helium looks like an $\alpha$ particle with shielded electric
charge. It can closely approach nuclei due to the absence of a
Coulomb barrier. On that reason it seems that in the presence of
OLe-helium the character of SBBN processes should change
drastically. However, this change might be not so dramatic
\cite{anutium}.

In fact, the size of OLe-helium is of the order of the size of
$^4He$ and for a nucleus $Z$ with electric charge $Z>2$ the size
of the Bohr orbit for an $A^{--} Z$ ion is less than the size of
nucleus $Z$. This means that while binding with a heavy nucleus
$A^{--}$ penetrates it and effectively interacts with a part of
the nucleus with a size less than the corresponding Bohr orbit.
This size corresponds to the size of $^4He$, making OLe-helium the
most bound $A^{--} Z$-atomic state. It favors the picture,
according to which OLe-helium collision with a nucleus, results in
the formation of OLe-helium and the whole process looks like an
elastic collision.

The interaction of the $^4He$ component of $(^4He^{++}A^{--})$ with a
$^A_ZQ$ nucleus can lead to a nuclear transformation due to the
reaction \beq ^A_ZQ+(^4He^{++}A^{--}) \rightarrow ^{A+4}_{Z+2}Q
+A^{--}, \label{EHeAZ} \eeq provided that the masses of the
initial and final nuclei satisfy the energy condition \beq M(A,Z)
+ M(4,2) - I_{OHe}> M(A+4,Z+2), \label{MEHeAZ} \eeq where $I_{OHe}
= 1.6 \MeV$ is the binding energy of OLe-helium and $M(4,2)$ is
the mass of the helium nucleus.

The final nucleus is formed in the excited $[\alpha, M(A,Z)]$
state, which can rapidly experience an $\alpha$-decay, giving rise to
a $(OHe)$ regeneration and to an effective quasi-elastic process of
$(OHe)$-nucleus scattering. It leads to a possible suppression of
the nuclear transformation (\ref{EHeAZ}).

The condition (\ref{MEHeAZ}) is not valid for stable nuclei
participating in reactions of the SBBN. However, unstable tritium
$^3H$, produced in SBBN and surviving 12.3 years after it, can
react with OLe-helium, forming $^7Li$ in process $^3H+(^4He
A^{--}) \rightarrow ^{7}Li +A^{--}$. The Alexium $A^{--}$, released in
this process, is captured by $^4He$ and regenerates OLe-helium,
while $^7Li$ reacts with OLe-helium, forming $^{11}B$ etc. After
$^{39}K$ the chain of transformations starts to create unstable
isotopes and gives rise to an extensive tree of transitions along
the table of nuclides. This set of processes involves the fraction
of baryons of the order of SBBN tritium abundance ($^3H/H \sim
10^{-7}$) and since it does not stop on lithium, but goes further
to nuclides which are observed now with much higher abundance, it
can not be excluded by a simple argument. This picture opens a new
path of chemical evolution of matter on the pre-galactic stage and
needs self-consistent consideration within a complete network of
nuclear processes.

\subsection{\label{OleHecat} OLe-helium catalysis of (AC) binding}
The process of $C^{++}$ capture by the $(OHe)$ atom looks as
follows \cite{leptons}. Being in thermal equilibrium with the
plasma, free $C^{++}$ have momentum $k = \sqrt{2T m_C}.$ If their
wavelength is much smaller than the size of the $(OHe)$ atom, they
can penetrate the atom and bind with $A^{--}$, expelling $He$ from
it. The rate of this process is determined by the size of the
$(OHe)$ atoms  and is given in \cite{leptons} as $$\sv_0 \sim \pi
R_{OHe} ^2 \sim \frac{\pi}{(\bar \alpha m_{He})^2} = \frac{\pi}{2
I_{OHe} m_{He}} \approx 3 \cdot 10^{-15}\frac{cm^3}{s}. $$ Here
$\bar \alpha = Z_A Z_{He} \alpha.$ At $T< T_a = \bar \alpha^2
m_{He} \frac{m_{He}}{2m_{C}} = \frac{I_{OHe}m_{He}}{m_{C}} = 4
\cdot 10^{-2} I_{OHe}/S_2$ the wavelength of $C^{++}$, $\lambda$,
exceeds the size of $(OHe)$ and the rate of $(OHe)$ catalysis is
suppressed by a factor $(R_{OHe}/\lambda)^3 = (T/T_a)^{3/2}$ and
is given by $$\sv_{cat}(T<T_a)= \sv_0 \cdot (T/T_a)^{3/2}.
$$

In the presence of the $y$-interaction both OLe-helium and $C^{++}$
are y-charged and for slow charged particles a Coulomb-like factor
of the "Sakharov enhancement" \cite{Sakhenhance} should be added in
these expressions
$$C_y=\frac{2 \pi \alpha_y/v}{1 - \exp{(-2 \pi \alpha_y/v)}},$$
where $v=\sqrt{2T/m}$ is relative velocity. It results in
\beq
\sv_0 = \pi R_{OHe} ^2 \cdot 2 \pi\alpha_y \cdot (m/2T)^{1/2} =
\frac{\alpha_y}{\bar \alpha} \frac{\pi^2}{I_{OHe} m_{He}} \cdot
(\frac{m}{ m_{He}})^{1/2}\frac{1}{x^{1/2}} \approx
10^{-13}\frac{\alpha_y}{1/30}(\frac{S_2}{x})^{1/2}\frac{cm^3}{s},
\label{Epsv} \eeq where $x=T/I_{OHe}.$ At $T < T_a$ the rate of
OLe-helium catalysis is given by \beq \sv_{cat}(T<T_a)= \sv_0
\cdot (T/T_a)^{3/2}=\frac{\alpha_y}{\bar \alpha} \frac{\pi^2}{T_a
m_{He}} \cdot (\frac{T}{T_a})\approx 2 \cdot 10^{-19}
\frac{\alpha_y}{1/30}S_2^2 (\frac{T}{300K})\frac{cm^3}{s}.
\label{catETa}
\eeq

The "Coulomb-like" attraction of $y$-charges can lead to their
radiative recombination. It can be described in the analogy to the
process of free monopole-antimonopole annihilation considered in
\cite{ZK}. The potential energy of the Coulomb-like interaction between
$A$ and $C$ exceeds their thermal energy $T$ at the distance
$$ d_0 = \frac{\alpha_y}{T}.$$
Following the classical solution of energy loss due to radiation,
converting infinite motion to finite, free $y$-charges form bound
systems at the impact parameter \cite{ZK,4had}
\beq a \approx
(T/m)^{3/10} \cdot d_0. \label{impact}
\eeq
The rate of such a
binding is then given by \cite{4had}
\beq \sv = \pi a^2 v \approx
\pi \cdot (m/T)^{9/10} \cdot (\frac{\alpha_y}{m})^2 \approx
\label{sigimpact}
\eeq
$$\approx 2 \cdot 10^{-12}
(\frac{\alpha_y}{1/30})^{2}(\frac{300K}{T})^{9/10}S_2^{-11/10}
\frac{cm^3}{s}$$
The successive evolution of this highly excited atom-like bound
system is determined by the loss of angular momentum owing to
the $y$-radiation. The time scale for the fall into the center in this
bound system, resulting in $AC$ recombination, was estimated
according to classical formula in \cite{DFK,4had}
\beq
\tau = \frac{a^3}{64 \pi} \cdot (\frac{m}{\alpha_y})^2 =
\frac{\alpha_y}{64 \pi} \cdot (\frac{m}{T})^{21/10} \cdot
\frac{1}{m} \label{recomb}
\eeq
$$\approx 2 \cdot 10^{-4} (\frac{\alpha_y}{1/30})(\frac{300K}{T})^{21/10}S_2^{11/10} s.$$
As is easily seen from Eq.(\ref{recomb}) this time scale of
$AC$ recombination $\tau \ll m/T^2 \ll m_{Pl}/T^2$ turns out to be
much less than the cosmological time at which the bound system was
formed.

The above classical description assumes $a=
\alpha_y/(m^{3/10}T^{7/10}) \gg  1/(\alpha_y m)$ and is valid at
$T \ll T_{rc}=m  \alpha_y^{20/7} \approx 60 \MeV
S_2(\frac{\alpha_y}{1/30})^{20/7}$ \cite{4had,Fargion:2005xz}.
Since $T_{rc} \gg I_{OHe}$ effects of radiative recombination can
also contribute $AC$-binding due to OLe-helium catalysis. However,
the rate of this binding is dominated by (\ref{catETa}) at
\beq T
\le T_a (\frac{\alpha_y \bar
\alpha^{6/5}}{\pi^2})^{10/19}(\frac{m_{He}}{m})^{11/19} \approx
100 \eV (\frac{\alpha_y}{1/30})^{10/19}S_2^{-30/19}
\label{sigrc}
\eeq
and the radiative recombination becomes important
only at a temperature much less, than in the period of cosmological
OLe-helium catalysis.

At modest values of $S_2$ the abundance of primordial
antiparticles is suppressed \cite{leptons} and the abundance of
free $C$, $r_C$, is equal to the abundance of $A^{--}$, trapped in
the $(OHe)$ atoms, $r_A=r_{OHe}$. Therefore a decrease of their
concentration due to the OLe-helium catalysis of $(AC)$ binding is
determined by the equation \beq
\frac{dr_{C}}{dx}=f_{1He}\left<\sigma v\right> r_{C}r_{OHe},
\label{Esrecomb} \eeq where $x=T/I_{OHe}$, $r_{OHe}=r_C$, $\sv$ is
given by Eqs.(\ref{Epsv}) at $T>T_a$ and (\ref{catETa}) at
$T<T_a$, $\bar \alpha = Z_A Z_{He} \alpha$ and
$$f_{1He}=\sqrt{\frac{\pi g_s^2}{45g_{\epsilon}}}m_{Pl}I_{OHe} \approx m_{Pl}I_{OHe}.$$
The solution of Eq.(\ref{Esrecomb}) is given by
$$r_C=r_{OHe}=\frac{r_{C0}}{1+r_{C0}J_{OHe}} \approx \frac{1}{J_{OHe}} \approx
7 \cdot 10^{-20}(\frac{1/30}{\alpha_y})/f(S_2).$$ Here
$$J_{OHe}=\int_0^{x_{fHe}} f_{1He}\left<\sigma v\right>dx =$$ \beq
= \pi^2 \frac{\alpha_y}{\bar \alpha}(\frac{ m_{Pl}}{2 m_{He}})
f(S_2) \approx 1.4 \cdot 10^{19}(\frac{\alpha_y}{1/30})\cdot
f(S_2), \label{JEHe} \eeq $x_{fHe}=1/27$ and the dependence on
$S_2$ is described by the function $f(S_2) =
4(\frac{S_2}{1.08})^{1/2} - 3$ for $S_2> 1.08$; $f(S_2) =
(\frac{S_2}{1.08})^{2}$ for $S_2 < 1.08.$

At large $S_2 > 40$ the primordial abundance of antiparticles
($\bar A$ and $\bar C$) is not suppressed. OLe-helium catalyze in
this case annihilation of these antiparticles through the
formation of AC-positronium and it was shown in \cite{leptons}
that electromagnetic showers, induced by annihilation products can
neither influence the light element abundance, nor cause
observable distortions of the CMB spectrum.

Colexium $C^{++}$ ions, which remain free after OLe-helium catalysis,
are in thermal equilibrium due to their Coulomb scattering with matter
plasma.
At $T < T_{od} \approx 1 \keV$ energy and momentum transfer due to nuclear interaction
from baryons to OLe-helium is not effective $n_b \sv (m_p/m_o) t < 1$.
Here \beq \sigma \approx \sigma_{OHe} \sim \pi R_{OHe}^2 \approx
10^{-25}\cm^2.\label{sigOHe} \eeq and $v = \sqrt{2T/m_p}$ is
baryon thermal velocity. Then OLe-helium gas decouples from plasma
and radiation and must behave like a sparse component of dark matter.
However, for a small window of parameters $1 \le S_2 \le 2$ at
$T<(\frac{\alpha_y}{1/30})\frac{10 \eV}{S_2 f(S_2)^2}$ Coulomb-like
scattering due to $y$ interaction with $C^{++}$ ions returns OLe-helium
to thermal equilibrium with plasma and supports effective energy and momentum
exchange between $A$ and $C$ components during all the pre-galactic stage.

5) $t \sim 2.5 \cdot 10^{11}\s$  at $T \sim I_{He}/30 \approx 2
eV.$ Here $I_{He}= Z^2 \alpha^2 m_{e}/2 = 54.4 \eV$ is the
potential energy of an ordinary He atom.  Free $C^{++}$ with
charge $Z=+2$ recombine with $e^-$ and form anomalous helium atoms
$(eeC^{++})$.

6) $t \sim 10^{12}\s$  at $T \sim T_{RM} \approx 1 \eV.$ AC-matter
dominance starts with $(AC)$-"atoms", playing the role of CDM in
the formation of Large Scale structures.

7) $z \sim 20.$ The formation of galaxies starts, triggering
$(AC)$ recombination in dense matter bodies.

\subsection{\label{Galaxy} Galaxy formation in the AC-Universe}
The development of gravitational instabilities of AC-atomic gas
follows the general path of the CDM scenario, but the composite
nature of $(AC)$-atoms leads to some specific difference. In
particular, one might expect that particles with a mass $m_{AC} =
200 S_2 \GeV$
 should form gravitationally bound objects
with the minimal mass \beq M = m_{Pl} (\frac{m_{Pl}}{m_{AC}})^2
\approx 5\cdot 10^{28}/S_2^2 \g. \label{MHEm} \eeq However, this
estimation is not valid for  composite CDM particles, which
$(AC)$-atoms are.

For $S_2<6$ the bulk of $(AC)$ bound states appear in the Universe
at $T_{fAC} = 0.7 S_2 \MeV$ and the minimal mass of their
gravitationally bound systems is given by the total mass of $(AC)$
within the cosmological horizon in this period, which is of the
order of \beq M = \frac{T_{RM}}{T_{fAC}} m_{Pl}
(\frac{m_{Pl}}{T_{fAC}})^2 \approx 6\cdot 10^{33}/S_2^3 \g,
\label{MHEm} \eeq where $T_{RM}=1 \eV$ corresponds to the
beginning of the AC-matter dominated stage.

If these objects, containing $N = 2 \cdot 10^{55} \cdot S_2^{-4}$
 $(AC)$-atoms, separated by the cosmological expansion at $z_s \sim
20$, they have an internal number density
$$n \approx 6 \cdot 10^{-5}\cdot S_2^{-1} (\frac{1+z_s}{1+20})^3 \cm^{-3}$$
and the size \beq R = (\frac{N}{4 \pi n/3})^{1/3}\approx 3 \cdot
10^{19}\cdot S_2^{-1} (\frac{1+20}{1+z_s}) \cm. \label{RHEgb} \eeq
At $S_2>6$ the bulk of $(AC)$-atoms is formed only at $T_{OHe} =
60 \keV$ due to OLe-helium catalysis. Therefore at $S_2>6$ the
minimal mass is independent of $S_2$ and is given by \beq M =
\frac{T_{RM}}{T_{OHe}} m_{Pl} (\frac{m_{Pl}}{T_{OHe}})^2 \approx
10^{37} \g. \label{MEPm} \eeq
The size of $(AC)$-"atoms" is ($Z_A=Z_C=2$) $$R_{AC} \sim 1/(Z_A
Z_C \alpha k_y m) \sim 1.37 \cdot 10^{-14}\cdot S_2^{-1}k_y^{-1}
\cm$$ and their mutual collision cross section is about \beq
\sigma_{AC} \sim \pi R_{AC}^2 \approx 6 \cdot 10^{-28}\cdot
S_2^{-2}k_y^{-2}\cm^2. \label{sigHEHE} \eeq AC-"atoms" can be
considered as collision-less gas in clouds with a number density
$n_{AC}$ and size $R$, if $n_{AC}R < 1/\sigma_{AC}$.

At small energy transfer $\Delta E \ll m$ cross section for
interaction of AC-atoms with matter is suppressed by the factor
$\sim Z^2 (\Delta E/m)^2$, being for scattering on nuclei with
charge $Z$ and atomic weight $A$ of the order of $\sigma_{ACZ}
\sim Z^2/\pi (\Delta E/m)^2 \sigma_{AC} \sim Z^2 A^2 10^{-43}
\cm^2 /S^2_2.$ Here we take $\Delta E \sim 2 A m_p v^2$ and $v/c
\sim 10^{-3}$ and find that even for heavy nuclei with $Z \sim
100$ and $A \sim 200$ this cross section does not exceed $4 \cdot
10^{-35} \cm^2 /S^2_2.$ It proves WIMP-like behavior of AC-atoms
in the ordinary matter.

The products of incomplete $AC$ binding - OLe-helium and anomalous
helium - have much stronger interaction with matter (nuclear and
atomic) and need special strategy for their direct experimental
search. In particular it should be noted that OLe-helium
represents a tiny fraction of dark matter and thus escapes severe
constraints \cite{McGuire:2001qj} on strongly interacting dark
matter particles (SIMPs) \cite{Starkman,McGuire:2001qj} imposed by
the XQC experiment \cite{XQC}.

Mutual collisions of AC-"atoms" determine the evolution timescale
for a gravitationally bound system of collision-less AC-gas \beq
t_{ev} = \frac{1}{n \sigma_{AC} v} \approx 10^{23} S_2^{17/6}
(\frac{1 \cm^{-3}}{n})^{7/6}\s, \label{evHEcoll} \eeq where the
relative velocity $v = \sqrt{G M/R}$ is taken at $S_2<6$ for a
cloud of mass Eq.(\ref{MHEm}) and an internal number density $n$.
The timescale Eq.(\ref{evHEcoll}) exceeds substantially the age of
the Universe even at $S_2<6$. Therefore the internal evolution of
AC-atomic self-gravitating clouds cannot lead to the formation of
dense objects.

\subsection{\label{Matter} Solution for the problem of Anomalous Helium}
The main possible danger for the considered cosmological scenario
is the over-production of primordial anomalous isotopes.
Pre-galactic abundance of anomalous helium (of $C$-lepton atoms
$(eeC^{++})$) exceeds by up to 10 orders of magnitude the
experimental upper limits on its content in terrestrial matter.
The only way to solve the problem of anomalous isotopes is to find
a possible reason for their low abundance inside the Earth and a
solution to this problem implies a mechanism of effective
suppression of anomalous helium in dense matter bodies (in
particular, inside the Earth).
The idea of such suppression, first proposed in \cite{fractons}
and recently realized in \cite{4had} is as follows \cite{leptons}.

If anomalous species have an initial abundance relative to baryons
$\xi_{i0}$, their recombination with the rate $\sv$ in a body with
baryonic number density $n$ reduces their abundance during the age
of the body $t_b$ down to \beq \xi_{i} = \frac{\xi_{i0}}{1 +
\xi_{i0}n \sv t_{b}}. \label{matsym} \eeq If $\xi_{i0} \gg 1/(n
\sv t_{b})$ in the result, the abundance is suppressed down to
\beq \xi_{i} = \frac{1}{n \sv t_{b}}. \label{supsym} \eeq To apply
this idea to the case of the AC-model, OLe-helium catalysis can be
considered as the mechanism of anomalous isotope suppression.

The mechanism of the above mentioned kind can not in principle
suppress the abundance of remnants in interstellar gas by more
than factor $f_g \sim 10^{-2}$, since at least $1\%$ of this gas
has never passed through stars or dense regions, in which such
mechanisms are viable. It may lead to the presence of a $C^{++}$
(and $A$) component in cosmic rays at a level $\sim f_g \xi_i$.
Therefore based on the AC-model one can expect the anomalous
helium and "antihelium" fractions in cosmic rays
\begin{eqnarray}
 \frac{C^{++}}{He} \le 10^{-10}/f(S_2),\nonumber\\
\frac{A^{--}}{He} \le 10^{-10}/f(S_2).
\end{eqnarray}
These predictions are hardly within the reach for future cosmic
ray experiments
even for $S_2 \sim 1$ and decrease with $S_2$ as $\propto
S_2^{-2}$ for $S_2 < 1.08$ and as $\propto
S_2^{-1/2}$ for $S_2 > 1.08$.

The crucial role of the $y$-attraction comes into the realization of the
above mentioned mechanism. The condition of $y$-charge neutrality
makes Ole-helium to follow anomalous helium atoms $(C^{++}e^-e^-)$
in their condensation in ordinary matter objects. Due to this
condition OLe-helium and anomalous helium can not separate and
$AC$ recombination goes on much more effectively, since its rate
is given now by (\ref{sigimpact})$$\sv \approx 2 \cdot 10^{-12}
(\frac{\alpha_y}{1/30})^{2}(\frac{300K}{T})^{9/10}S_2^{-11/10}
\frac{cm^3}{s}. $$ This increase of recombination rate reduces
primeval anomalous helium (and OLe-helium) terrestrial content
down to $r \le 5 \cdot 10^{-30}.$

In the framework of our consideration, interstellar gas contains a
significant ($\sim f_g \xi_{C}$) fraction of $(eeC^{++})$. When
the interstellar gas approaches Solar System, it might be stopped
by the Solar wind in the heliopause at a distance $R_h \sim
10^{15} \cm$ from the Sun. In the absence of detailed experimental
information about the properties of this region we can assume for
our estimation following \cite{leptons} that all the incoming
ordinary interstellar gas, containing dominantly ordinary
hydrogen, is concentrated in the heliopause and the fraction of
this gas, penetrating this region towards the Sun, is pushed back
to the heliopause by the Solar wind. In the result, to the present
time during the age of the Solar system $t_E$ a semisphere of
width $L\sim R_h$ is formed in the distance $R_h$, being filled by
a gas with density $n_{hel} \sim (2 \pi R_h^2 v_g t_E n_g)/(2 \pi
R_h^2 L) \sim 10^8 \cm^{-3}.$ The above estimations show that this
region is transparent for $(OHe)$, but opaque for atomic size
remnants, in particular, for $(eeC^{++})$. Owing to the
$y$-interaction both components can thus be stopped in heliopause.
Though the Solar wind cannot directly stop heavy $(eeC^{++})$, the
gas shield in the heliopause slows down their income to Earth and
suppresses the incoming flux $I_{C}$ by a factor $S_h \sim
1/(n_{hel} R_h \sigma_{tra})$, where  $\sigma_{tra} \approx
10^{-18} S_2 \cm^2$. So the incoming flux, reaching the Earth, can
be estimated as \cite{4had,leptons} \beq I_{C} = \frac{\xi_{C} f_g
n_g v_g}{8\pi}S_h \approx \frac{10^{-10}}{f(S_2)}\frac{S_h}{5
\cdot 10^{-5}}(cm^2 \cdot s \cdot ster)^{-1}. \label{IincP} \eeq
Here $n_g \sim 1 \cm^{-3}$ and $v_g \sim 2 \cdot 10^{4} \cm/\s.$

Kinetic equilibrium between interstellar AC-gas pollution and $AC$
recombination in Earth holds \cite{4had} their concentration in
terrestrial matter at the level \beq n = \sqrt{\frac{j}{\sv}},
\label{statsol} \eeq where \beq j_{A}=j_{C}=j \sim \frac{2 \pi
I_C}{L} = 2.5 \cdot 10^{-11} S_h/f(S_2) \cm^{-3}\s^{-1},
\label{statin} \eeq within the water-circulating surface layer of
thickness $L \approx 4 \cdot 10^5 \cm$.  Here $I_{C} \approx 2
\cdot 10^{-6} S_h (\cm^2 \cdot \s \cdot ster)^{-1}$ is given by
Eq.(\ref{IincP}), factor $S_h$ of incoming flux suppression in
heliopause can be as small as $S_h \approx 5 \cdot 10^{-5}$ and
$\sv$ is given by the Eq.(\ref{sigimpact}). For these values of
$j$ and $\sv$ one obtains in water \beq n \le 3.5
\sqrt{S_h/f(S_2)} (\frac{1/30}{\alpha_y})S_2^{11/20}
\cm^{-3}.\label{ncPE} \eeq It corresponds to a terrestrial anomalous
helium abundance
$$ r \le 3.5 \cdot 10^{-23}\sqrt{S_h/f(S_2)}(\frac{1/30}{\alpha_y})S_2^{11/20},$$
being below the above mentioned experimental upper limits for
anomalous helium ($ r < 10^{-19}$).

The reduction of the anomalous helium abundance due to $AC$ recombination
in dense matter objects is not accompanied by an annihilation, which
was the case for $U$-hadrons in \cite{4had}, therefore the AC-model
escapes the severe constraints \cite{4had} on the products of such an
annihilation, which follow from the observed gamma background and
the data on neutrino and upward muon fluxes.

\subsection{\label{EpMeffects} Effects of $(OHe)$ catalyzed processes in the Earth}
The first evident consequence of the proposed excess is the
inevitable presence of $(OHe)$ in terrestrial matter. $(OHe)$
concentration in the Earth can reach the value (\ref{ncPE}) for
the incoming $(OHe)$ flux, given by Eq.(\ref{IincP}). The
relatively small size of neutral $(OHe)$ may provide a catalysis
of cold nuclear reactions in ordinary matter (much more
effectively, than muon catalysis). This effect needs special and
thorough nuclear physical treatment. On the other hand, if
$A^{--}$ capture by nuclei, heavier than helium, is not effective
and does not lead to a copious production of anomalous isotopes,
$(OHe)$ diffusion in matter is determined by an elastic collision
cross section (\ref{sigpEpcap}) and may effectively hide
OLe-helium from observations.

One can give the following argument for an effective regeneration
of OLe-helium in terrestrial matter. OLe-helium can be destroyed
in reactions (\ref{EHeAZ}). Then free $A^{--}$ are released and
owing to a hybrid Auger effect (capture of $A^{--}$ and ejection
of ordinary $e$ from the atom with atomic number $A$ and charge of
$Z$ of the nucleus) $A^{--}$-atoms are formed, in which $A^{--}$
occupies highly an excited level of the $(Z-A^{--})$ system, which
is still much deeper than the lowest electronic shell of the
considered atom. $(Z-A^{--})$-atomic transitions to lower-lying
states cause radiation in the range intermediate between atomic
and nuclear transitions. In course of this falling down to the
center of the $(Z-A^{--})$ system, the nucleus approaches
$A^{--}$. For $A>3$ the energy of the lowest state $n$ (given by
$E_n=\frac{M \bar \alpha^2}{2 n^2} = \frac{2 A m_p Z^2
\alpha^2}{n^2}$)  of the $(Z-A^{--})$ system (having reduced mass
$M \approx A m_p$) with a Bohr orbit, $r_n=\frac{n}{M \bar
\alpha}= \frac{n}{2 A Z m_p \alpha}$, exceeding the size of the
nucleus, $r_A \sim A^{1/3}m^{-1}_{\pi}$, is less, than the binding
energy of $(OHe)$. Therefore the regeneration of OLe-helium in a
reaction, inverse to (\ref{EHeAZ}), takes place. An additional
reason for dominantly elastic channel for reactions (\ref{EHeAZ})
is that the final state nucleus is created in the excited state
and its de-excitation via $\alpha$-decay can also result in
OLe-helium regeneration.

Another effect is the energy release from OLe-helium catalysis of
$(AC)$ binding. The consequences of the latter process are not as
pronounced as those discussed in \cite{4had,leptons} for the
annihilation of  4th generation hadrons in terrestrial matter, but
it might lead to a possible test for the considered model.

In our mechanism the terrestrial abundance of anomalous $(C^{++}ee)$
is suppressed due to the $(OHe)$ catalyzed binding of most of the $C$
from the incoming flux $I_{C}$, reaching the Earth. $AC$ binding
is accompanied by de-excitation of the initially formed bound
$(AC)$ state. To expel $^4He$ from OLe-helium, this state should
have binding energy exceeding $I_{He}=1.6 MeV$, therefore MeV
range $\gamma$ transitions from the lowest excited levels to the
ground state of $(AC)$ with $I_{AC}=80 S_2 k_y MeV$ should take
place. The danger of gamma radiation from these transitions is
determined by the actual magnitude of the incoming flux, which was
estimated in subsection \ref{Matter} as Eq.(\ref{IincP}).

The stationary regime of $(OHe)$ catalyzed recombination of these
incoming $C^{++}$ in the Earth should be accompanied by gamma
radiation with the flux $F(E) = N(E) I_C l_{\gamma}/R_E$, where
$N(E)$ is the energy dependence of the multiplicity  of $\gamma$
quanta with energy $E$ in $(AC)$-atomic transitions, $R_E$ is the
radius of the Earth and $l_{\gamma}$ is the mean free path of such
$\gamma$ quanta. At $E
> 10 MeV$ one can roughly estimate the flux $F(E> 10 MeV) \sim
\cdot \frac{10^{-16}}{f(S_2)}\frac{S_h}{5 \cdot 10^{-5}} (cm^2
\cdot s \cdot ster)^{-1}$, coming from the atmosphere and the
surface layer $l_{\gamma} \sim 10^3 cm$. Even without the
heliopause suppression (namely, taking $S_h = 1$) $\gamma$
radiation from $AC$ binding seems to be hardly detectable.

In the course of $(AC)$ atom formation electromagnetic transitions
with $\Delta E > 1 \MeV$ can be a source of $e^+e^-$ pairs, either
directly with probability $\sim 10^{-2}$ or due to development of
electromagnetic cascade. If $AC$ recombination goes on
homogeneously in Earth within the water-circulating surface layer
of the depth $L \sim 4 \cdot 10^5 \cm$ inside the volume of Super
Kamiokande with size $l_{K} \sim 3 \cdot 10^3 \cm$ equilibrium
$AC$ recombination should result in a flux of $e^+e^-$ pairs $F_e
= N_e I_C l_{K}/L$, which for $N_e \sim 1$ can be as large as $F_e
\sim \cdot \frac{10^{-12}}{f(S_2)}\frac{S_h}{5 \cdot 10^{-5}}
(cm^2 \cdot s \cdot ster)^{-1}.$

Such an internal source of electromagnetic showers in large volume
detectors inevitably accompanies the reduction of the anomalous helium
abundance due to $AC$ recombination and might give an advantage of
experimental tests for the considered model. Their signal might be
easily  disentangled \cite{leptons} (above a few MeV range) with respect
to common charged
    current neutrino interactions and single electron tracks,
     because the tens MeV gamma lead, by pair productions, to twin
    electron  tracks,  nearly aligned along their Cerenkov rings.
    The signal is piling  the energy in windows where
   few atmospheric neutrino and  cosmic Super-Novae  radiate. The same gamma flux produced is comparable
    to expected secondaries of tau decay secondaries  while showering  in air at
    the horizons edges(\cite{FarSolar03},\cite{FarTau02},\cite{FarTau03}). The predicted signal
strongly depends, however, on the uncertain astrophysical
parameters (concentration OLe-helium and anomalous helium in the
interstellar gas, their flux coming to Earth etc) as well as on
the geophysical details of the actual distribution of OLe-helium and
anomalous helium in the terrestrial matter, surrounding large
volume detectors.

Direct search for OLe-helium from dark matter halo is not possible in
underground detectors due to OLe-helium slowing down in terrestrial matter.
Therefore special strategy of such search is needed, which can exploit
sensitive dark matter detectors before they are installed under ground.
In particular, future superfluid $^3He$ detector \cite{Winkelmann:2005un}
and even its existing few $\g$ laboratory prototype can be used to
put constraints on the in-falling OLe-helium flux from galactic halo.

\subsection{\label{EpMeffects} $(OHe)$ catalyzed formation of $(AC)$-matter
objects inside ordinary matter stars and planets}

$(AC)$-atoms  from the halo interact weakly with ordinary matter
and can be hardly captured in large amounts by a matter object.
However the following mechanism can provide the existence of a
significant amount of $(AC)$-atoms in matter bodies and even the
formation of gravitationally bound dense $(AC)$- bodies inside
them.

Inside a dense matter body $(OHe)$ catalyzes $C$ aggregation into
$(AC)$-atoms in the reaction \beq (eeC^{++}) + (A^{--}He)
\rightarrow (AC)+He+2e. \label{EpUUUEe} \eeq In the result of this
reaction $(OHe)$, interacting with matter with a nuclear cross
section given by \beq \sigma_{trAb} =  \pi R_{OHe}^2
\frac{m_{p}}{m_A} \approx 10^{-27}/S_2 \cm^{2}, \label{sigpEpcap}
\eeq and $(eeC^{++})$, having a nearly atomic cross section of
that interaction \beq \sigma_{tra}= \sigma_a (m_p/m_C) \approx
10^{-18} S_2 \cm^2, \label{sigpUEcap} \eeq bind into weakly
interacting $(AC)$-atom, which decouples from the surrounding
matter.

In this process "products of incomplete AC-matter combustion"
(OLe-helium and anomalous helium), which were coupled to the
ordinary matter by hadronic and atomic interactions, convert into
$(AC)$ atoms, which immediately sinks down to the center of the
body.

The  amount of $(AC)$-atoms produced inside matter object by the
above mechanism is determined by the initial concentrations of
OLe-helium $(A^{--}He)$ and anomalous helium atoms $(eeC^{++})$.
This amount $N$ defines the number density of $(AC)$-matter inside
the object, being initially $n \sim N/R_s^3$, where $R_s$ is the
size of body. At the collision timescale $t \sim (n \sigma_{AC}
R_s)^{-1}$, where the $(AC)$-atom collision cross section is given
by Eq.(\ref{sigHEHE}), in the central part of body a dense and
opaque $(AC)$-atomic core is formed. This core is surrounded by a
cloud of free $(AC)$-atoms, distributed as $\propto R^{-2}$.
Growth and evolution of this $(AC)$-atomic conglomeration may lead
to the formation of a dense self-gravitating $(AC)$-matter object,
which can survive after the star, inside which it was formed,
exploded.

The relatively small mass fraction of AC-matter inside matter
bodies corresponds to the mass of the $(AC)$-atomic core $\ll
10^{-4} S_2 M_{\odot}$ and this mass of AC-matter can be hardly
put within its gravitational radius in the result  of the
$(AC)$-atomic core evolution. Therefore it is highly improbable
that such an evolution could lead to the formation of black holes
inside matter bodies.

\section{\label{Discussion} Discussion}
In the present paper we explored the cosmological implications
of the AC-model presented in \cite{5,leptons} with an additional Coulomb
like interaction, mediated by the $y$-photon. This new $U(1)$ interaction
appears naturally in the almost-commutative framework. For
the standard model particles the $y$-photons are invisible,
the only source of this invisible light are the AC-particles.
Due to this new strict gauge symmetry the AC-leptons
acquire stability, similar to the case of 4th generation hadrons \cite{4had}
and fractons \cite{fractons}.

The AC-particles are lepton like, coupling apart from  the
$y$-photons only to the ordinary photon and the $Z$-boson. Their
electric charge is taken to be $-2e$ for the $A^{--}$-lepton and
$+2e$ for the $C^{++}$-lepton. They may form atom like bound
states $(A^{--}C^{++})$ with WIMP like cross section which can
play the role of evanescent Cold Dark Matter in the modern
Universe. The AC-model escapes most of the drastic problem of the
Sinister Universe \cite{Glashow}, related with the primordial
$^4He$ cage for $-1$ charge particles and a consequent
overproduction of anomalous hydrogen \cite{Fargion:2005xz}. These
charged $^4He$ cages pose a serious problems to CDM models with
single charged particles, since their Coulomb barrier prevents
successful recombination of positively and negatively charged
particles. The doubly charged $A^{--}$-leptons bind with helium in
the neutral OLe-helium catalyzers of $AC$ binding and AC-leptons
may thus escape this trap.

Nonetheless the binding of AC-leptons into $(AC)$-"atoms" is a
multi step process, which, due to the expansion of the Universe,
produces necessarily exotic combinations of AC-matter and ordinary
matter, as well as free charged AC-ions. A mechanism to suppress
these unwanted remnants is given by the OLe helium catalysis
$(AHe)+C \rightarrow (AC)+He$. This process is enhanced by the
long-range interaction between the AC-leptons due to the
$y$-photons. It prevents the fractionating of  AC-particles and in
this way enhances also the binding of AC-particles in dense matter
bodies today. This process is necessary to efficiently suppress
exotic atoms to avoid the strong observational bounds.

The AC-model with $y$-interaction may thus solve the serious
problem of anomalous atoms, such as anomalous helium, which
appeared in the AC-cosmology presented in \cite{leptons} as well
as the question of the stability of the AC-leptons. However the
AC-cosmology, even with $y$-interaction, can only be viewed as  an
illustration of the possible solution for the Dark Matter problem
since the following problems remain open:

1. The reason for particle-antiparticle asymmetry.

The AC-model cannot provide a mechanism to explain the necessary
particle-antiparticle asymmetry. Such a mechanism may arise from
further extensions of the AC-model within noncommutative geometry
or due to phenomena from quantum gravity.

2. Possibly observable nuclear processes due to
OLe helium.

A challenging problem is the
possible existence of OLe helium $(AHe)$ and of nuclear
transformations, catalyzed by $(AHe)$. The question about its
consistency with observations remains open, since special nuclear
physics analysis is needed to reveal what are the actual
OLe-helium effects in SBBN and in terrestrial matter.

3. Recombination of AC-particles in dense matter objects.

  The recombination into (AC)-atoms and the consequent release
 of gamma energy at tens MeV, at the edge of detection in Super Kamiokande underground
  detector, (at rate comparable to cosmic neutrino Supernovae
  noise or Solar Flare thresholds \cite{FarSolar03}).
    Their signal might be easily  disentangled \cite{leptons} (above a few MeVs ) respect common charged
    current neutrino interactions and single electron tracks
     because the tens MeV gamma lead, by pair productions, to twin
    electron  tracks,  nearly aligned along their Cerenkov rings.
    The signal is piling  the energy in windows where
   few atmospheric neutrino and  cosmic Super-Novae  radiate.

4. Mixing of $y$-photons with neutral gauge bosons.

Due to the interaction of AC-leptons with photons and $Z$-bosons
the invisible $y$-photons will appear in fermionic AC-loops. Thus
standard model fermions may acquire a weak long-range
$y$-interaction. Furthermore it may be necessary to take the
AC-lepton loops into account for high precision calculations of
QED parameters such as the anomalous magnetic moment of the muon.
Since these parameters are extremely well known, they may provide
a crucial lower bound for the mass of the AC-particles.

In the context of AC-cosmology search for AC leptons at accelerators
acquires the meaning of crucial test for the existence of basic components
of the composite dark matter. One can hardly overestimate the
significance of positive results of such searches, if AC leptons
really exist and possess new long range interaction.

To conclude, in the presence of the $y$-interaction AC-cosmology can
naturally resolve the problem of anomalous helium, avoiding all
the observational constraints on the effects, accompanying
reduction of its concentration.
Therefore the AC-model with invisible light for its dark matter
components might provide a realistic model of composite dark
matter.

\paragraph*{{\bf Note added:}} This paper has been merged with
\cite{leptons} for publication in Class. Quantum Grav. \cite{CQG}.

\section*{Acknowledgments}
We are grateful to D.Fargion and T. Sch\"ucker for fruitful discussions.
C.A.S. is grateful to the Alexander von Humboldt-Stiftung.
M.Kh.
is grateful to LPSC, Grenoble, France for hospitality and to
D.Rouable for help.

\section*{Appendix 1: Basic definitions of noncommutative geometry}

In this section we will give the necessary basic definitions for a classification of almost commutative
geometries from a particle physics point of view. As mentioned above only the matrix part
will be taken into account, so we restrict ourselves to real, $S^0$-real, finite spectral triples
($\mathcal{A},\mathcal{H},\mathcal{D}, $ $J,\epsilon,\chi$). The algebra $\mathcal{A}$ is
a finite sum of matrix algebras
$\mathcal{A}= \oplus_{i=1}^{N} M_{n_i}(\mathbb{K}_i)$ with $\mathbb{K}_i=\mathbb{R},\mathbb{C},\mathbb{H}$ where $\mathbb{H}$
denotes the quaternions.
A faithful representation $\rho$ of $\mathcal{A}$ is given on the finite dimensional Hilbert space $\mathcal{H}$.
The Dirac operator $\mathcal{D}$ is a selfadjoint operator on $\mathcal{H}$ and plays the role of the fermionic mass matrix.
$J$ is an antiunitary involution, $J^2=1$, and is interpreted as the charge conjugation
operator of particle physics.
The $S^0$-real structure $\epsilon$ is a unitary involution, $\epsilon^2=1$. Its eigenstates with
eigenvalue $+1$ are the particle states, eigenvalue $-1$ indicates antiparticle states.
The chirality $\chi$ is as well a unitary involution, $\chi^2=1$, whose eigenstates with eigenvalue
$+1$ $(-1)$ are interpreted as right (left) particle states.
These operators are required to fulfill Connes' axioms for spectral triples:

\begin{itemize}
\item  $[J,\mathcal{D}]=[J,\chi]=[\epsilon,\chi]=[\epsilon,\mathcal{D}]=0, \quad \epsilon
J=-J \epsilon,\quad\mathcal{D}\chi =-\chi \mathcal{D}$,

$[\chi,\rho(a)]=[\epsilon,\rho(a)]=[\rho(a),J\rho(b)J^{-1}]=
[[\mathcal{D},\rho(a)],J\rho(b)J^{-1}]=0, \forall a,b \in \mathcal{A}$.
\item The chirality can be written as a finite sum $\chi =\sum_i\rho(a_i)J\rho(b_i)J^{-1}.$
This condition is called {\it orientability}.
\item The intersection form
$\cap_{ij}:=\T(\chi \,\rho (p_i) J \rho (p_j) J^{-1})$ is non-degenerate,
$\rm{det}\,\cap\not=0$. The
$p_i$ are minimal rank projections in $\mathcal{A}$. This condition is called
{\it Poincar\'e duality}.
\end{itemize}
Now the Hilbert space $\mathcal{H}$ and the representation $\rho$ decompose with respect to the
eigenvalues of $\epsilon$ and $\chi$ into left and right, particle and antiparticle spinors
and representations:
\begin{eqnarray}
\mathcal{H}=\mathcal{H}_L\oplus\mathcal{H}_R\oplus\mathcal{H}_L^c\oplus\mathcal{H}_R^c
\nonumber
\end{eqnarray}
\begin{eqnarray}
\rho = \rho_L \oplus \rho_R \oplus \overline{ \rho_L^c} \oplus \overline{ \rho_R^c}
\label{representation}
\end{eqnarray}
In this representation the Dirac operator has the form
\begin{eqnarray}
\mathcal{D}=\pp{0&\mathcal{M}&0&0\\
\mathcal{M}^*&0&0&0\\ 0&0&0&\overline{\mathcal{M}}\\
0&0&\overline{\mathcal{M}^*}&0}, \label{opdirac}
\end{eqnarray}
where $\mathcal{M}$ is the fermionic mass matrix connecting the left and the right handed Fermions.

Since the individual matrix algebras have only one irreducible representation for $\mathbb{K}=
\mathbb{R},\mathbb{H}$ and two for $\mathbb{K}=\mathbb{C}$ (the fundamental one and its complex
conjugate), $\rho$ may be written as a direct sum of these fundamental representations with
mulitiplicities
\begin{eqnarray}
\rho(\oplus_{i=1}^N a_i):=(\oplus_{i,j=1}^N
a_i \otimes
1_{m_{ji}} \otimes 1_{(n_j)})\
\oplus\ ( \oplus_{i,j=1}^N 1_{(n_i)} \otimes 1_{m_{ji}} \otimes
\overline{a_j} ).
\nonumber
\end{eqnarray}
There arise certain subtleties which are described in detail in
\cite{1,Kraj,pasch} and will be treated in a later extension of our work.

The first summand denotes the particle sector and the second the antiparticle sector. For the dimensions
of the unity matrices we have $(n)=n$ for $\mathbb{K}=\mathbb{R},\mathbb{C}$ and $(n)=2n$ for
$\mathbb{K}=\mathbb{H}$ and the convention $1_0=0$.
The multiplicities $m_{ji}$ are non-negative integers. Acting with the real structure
$J$ on $\rho$ permutes the main summands and complex conjugates them. It is also possible to write
the chirality as a direct sum
\begin{eqnarray}
\chi=(\oplus_{i,j=1}^N 1_{(n_i)} \otimes \chi_{ji}1_{m_{ji}} \otimes
1_{(n_j)})\
\oplus\
(\oplus_{i,j=1}^N 1_{(n_i)} \otimes \chi_{ji}1_{m_{ji}} \otimes 1_{(n_j)}),
\nonumber
\end{eqnarray}
where $\chi_{ji}=\pm 1$ according to our previous convention on left-(right-)handed spinors.
One can now define the multiplicity matrix $\mu \in M_N(\mathbb{Z})$ such that
$\mu _{ji}:=\chi _ {ji}\, m_{ji}$. This matrix is symmetric and decomposes into a particle and an antiparticle matrix, the second being just the particle matrix transposed, $\mu= \mu_P + \mu_A = \mu_P + \mu_P^T$. The intersection form of the Poincar\'e duality is now simply $\cap = \mu + \mu^T$, see \cite{Kraj,pasch}.

The mass matrix $\mathcal{M}$ of the Dirac operator connects the left and the right handed Fermions. Using
the decomposition of the representation $\rho$ and the corresponding decomposition of the Hilbert
space $\mathcal{H}$ we find two types of submatrices in $\mathcal{M}$, namely $M\otimes 1_{(n_k)}$ and
$1_{(n_k)}\otimes M$. $M$ is a complex $(n_i)\times(n_j)$ matrix connecting the i-th and the j-th sub-Hilbert
space and its eigenvalues give the masses of the fermion multiplet. We will call the k-th algebra the
colour algebra.

\section*{Appendix 2: Irreducibility, Non-Degeneracy}

To classify the almost commutative spectral triples we will impose
some extra conditions as in \cite{1}. We will require the
spectral triples to be irreducible and non-degenerate according to
the following definitions:

\begin{defn}
 i) A spectral triple $(\mathcal{A},\mathcal{H},\mathcal{D})$ is {\it degenerate} if the kernel of
$\mathcal{D}$ contains a non-trivial subspace of the complex Hilbert space $\mathcal{H}$
invariant under the representation $\rho$ on $\mathcal{H}$ of the real algebra
 $\mathcal{A}$.  \\
 ii) A non-degenerate spectral triple $(\mathcal{A},\mathcal{H},\mathcal{D})$ is {\it reducible} if
there is a proper subspace
$\mathcal{H}_0\subset\mathcal{H}$ invariant under the algebra $\rho(\mathcal{A})$  such that
$(\mathcal{A},\mathcal{H}_0,\mathcal{D}|_{\mathcal{H}_0})$ is a non-degenerate spectral triple. If the
triple is real, $S^0$-real and even, we require  the subspace
$\mathcal{H}_0$ to be also invariant under the real structure $J$, the $ S^0$-real
structure $\epsilon $ and under the chirality
$\chi $ such that the triple $(\mathcal{A},\mathcal{H}_0,\mathcal{D}|_{\mathcal{H}_0})$ is again real,
$S^0$-real and even.
\end{defn}

\begin{defn} The irreducible
 spectral triple $(\mathcal{A},\mathcal{H},\mathcal{D})$ is {\it dynamically non-degenerate} if all
minima $\ddfm$ of the action $V(\ddf)$ define a non-degenerate spectral
triple $(\mathcal{A},\mathcal{H},\ddfm )$ and if the spectra of all minima  have no
degeneracies other than the three kinematical degeneracies: left-right,
particle-antiparticle and colour. Of course in the massless case there is no
left-right degeneracy. We also suppose that the colour degeneracies are
protected by the little group. By this we mean that all eigenvectors of
$\ddfm$ corresponding to the same eigenvalue are in a common orbit of
the little group (and scalar multiplication and charge conjugation).
\label{irred}
\end{defn}

In physicists' language non-degeneracy excludes all models with
pairwise equal fermion masses in the left handed particle sector
up to colour degeneracy. Irreducibility means that we restrict
ourselves to one fermion generation and wish to keep the number of
fermions as small as allowed by the axioms for spectral triples.
The last requirement of definition \ref{irred} means
noncommutative colour groups are unbroken. It ensures that the
corresponding mass degeneracies are protected from quantum
corrections. It should be noted that the standard model of
particle physics meets all these requirements.

\section*{Appendix 3: Krajewski Diagrams}
Connes' axioms, the decomposition of the Hilbert space, the
representation and the Dirac operator allow a diagrammatic
dipiction. As was shown in \cite{Kraj} and \cite{1} this can be
boiled down to simple arrows, which encode the multiplicity matrix
and the fermionic mass matrix. From this information all the
ingredients of the spectral triple can be recovered. For our
purpose a simple arrow and connections of arrows at one point
(i.e. double arrows, edges, etc) are sufficient. The arrows always
point from right fermions (positive chirality) to left fermions
(negative chirality). We may also restrict ourselves to the
particle part, since the information of the antiparticle part is
included by transposing the particle part. We will adopt the
conventions of \cite{1} so that algebra elements tensorised with
$1_{m_{ij}}$ will be written as a direct sum of $m_{ij}$ summands.

$\bullet$ The Dirac operator: The components of the (internal) Dirac
operator are represented by horizontal or vertical lines connecting two
nonvanishing entries of opposite signs in the multiplicity matrix $\mu $
and we will orient them from plus to minus. Each arrow represents a
nonvanishing, complex submatrix in the Dirac operator: For instance
$\mu_{ij}$ can be linked to $\mu_{ik}$ or $\mu_{kj}$ by
\begin{center}\begin{tabular}{cc}
\rxy{
,(0,0)*\cir(0.7,0){}
,(5,0)*\cir(0.7,0){}
,(5,0);(0,0)**\dir{-}?(.6)*\dir{>}
,(0,-3)*{\mu_{ij}}
,(5,-3)*{\mu_{ik}}
}
&\;\;\;\;\;
\rxy{
,(0,0)*\cir(0.7,0){}
,(0,-5)*\cir(0.7,0){}
,(0,0);(0,-5)**\dir{-}?(.6)*\dir{>}
,(-3,0)*{\mu_{kj}}
,(-3,-5)*{\mu_{ij}}
}
\end{tabular} \end{center}
and these arrows represent respectively submatrices of $\mathcal{M}$ in $\mathcal{D}$ of
type $M\otimes 1_{(n_i)}$ with $M$ a complex $(n_j)\times(n_k)$ matrix
and $1_{(n_j)}\otimes M$  with $M$ a complex $(n_i)\times(n_k)$ matrix.

\noindent The requirement of
non-degeneracy of a spectral triple means that every nonvanishing
entry in the multiplicity matrix
$\mu $ is touched by at least one arrow.

$\bullet$ Convention for the diagrams: We will see that (for sums of up to
three simple algebras) irreducibility implies that most entries of $\mu $
have an absolute value less than or equal to two. So we will use a {\it simple
arrow} to connect plus one to minus one and {\it double arrows} to
connect plus one to minus two or plus two to minus one:)
\begin{center}
\begin{tabular}{ccc}
\rxy{
,(0,0)*\cir(0.7,0){}
,(5,0)*\cir(0.7,0){}
,(5,0);(0,0)**\dir{-}?(.6)*\dir{>}
,(0,-3)*{-1}
,(5,-3)*{+1}
}
&
\;\;\;\;\;
\rxy{
,(20,0)*\cir(0.7,0){}
,(20,0)*\cir(0.4,0){}*\frm{*}
,(15,0)*\cir(0.7,0){}
,(20,0);(15,0)**\dir2{-}?(.6)*\dir2{>}
,(20,-3)*{+1}
,(15,-3)*{-2}
}
&
\;\;\;\;\;
\rxy{
,(35,0)*\cir(0.7,0){}
,(30,0)*\cir(0.7,0){}
,(30,0)*\cir(0.4,0){}*\frm{*}
,(35,0);(30,0)**\dir2{-}?(.6)*\dir2{>}
,(35,-3)*{+2}
,(30,-3)*{-1}
}
\\
\\
\end{tabular}
\end{center}
Multiple arrows beginning or ending at one point are with or without edges are built
in an obvious way iterating the procedure above. We will give examples below that will
clarify these constructions.

Our arrows always point from plus, that is right chirality, to minus, that is
left chirality.
For a given algebra, every spectral triple is encoded in its
multiplicity matrix which itself is encoded in its Krajewski diagram, a field
of arrows. In our conventions, for particles, $\epsilon =1$, the column label
of the multiplicity matrix indicates the representation, the row label
indicates the multiplicity. For antiparticles, the row label
of the multiplicity matrix indicates the representation, the column label
indicates the multiplicity.

\noindent Every arrow comes with three algebras:
Two algebras that localize its end
points, let us call them {\it right and left algebras}
and a third algebra that localizes the arrow, let us call it {\it colour
algebra}.  For example for the arrow
\bb\rxy{
,(0,0)*\cir(0.7,0){}
,(5,0)*\cir(0.7,0){}
,(5,0);(0,0)**\dir{-}?(.6)*\dir{>}
,(0,-3)*{\mu _{ij}}
,(5,-3)*{\mu _{ik}}}\eee
the left algebra is $\mathcal{A} _j$, the right algebra is $\mathcal{A}_k$ and the colour
algebra is $\mathcal{A}_i$.

\noindent The {\it circles} in the diagrams only intend to guide the
eye. A {\it black disk} on a multiple arrow indicates that the coefficient of the
multiplicity matrix is plus or minus one at this location, ``the arrows
are joined at this location''.  For example the the following arrows
\begin{center}
\begin{tabular}{cc}
\rxy{
,(20,0)*\cir(0.7,0){}
,(20,0)*\cir(0.4,0){}*\frm{*}
,(15,0)*\cir(0.7,0){}
,(20,0);(15,0)**\dir2{-}?(.6)*\dir2{>}
,(20,-3)*{\mu_{ik}}
,(15,-3)*{\mu_{ij}}
}
&\quad
\rxy{
,(35,0)*\cir(0.7,0){}
,(30,0)*\cir(0.7,0){}
,(30,0)*\cir(0.4,0){}*\frm{*}
,(35,0);(30,0)**\dir2{-}?(.6)*\dir2{>}
,(35,-3)*{\mu_{ik}}
,(30,-3)*{\mu_{ij}}
}
\end{tabular}
\end{center}
\bb
\rxycc{0.7}{
,(5,-5)*\cir(0.4,0){}*\frm{*}
,(10,-5);(5,-5)**\dir{-}?(.6)*\dir{>}
,(5,-10);(5,-5)**\dir{-}?(.6)*\dir{>}
}
\eee
represent respectively submatrices of
$\mathcal{M}$ of type
$$\pp{M_1\cr M_2} \otimes 1_{(n_i)}\quad
{\rm and} \quad
\pp{M_1&M_2} \otimes 1_{(n_i)}
$$
with $M_1,M_2$ of size $(n_j)\times(n_k)$ or
in the third case, a matrix of type $\pp{M_1\otimes 1_{(n_i)}&1_{(n_j)}
\otimes M_2}$  where $M_1$ and $M_2$ are of size $(n_j)\times(n_k)$ and
$(n_i)\times(n_\ell)$.

\begin{figure}[h]
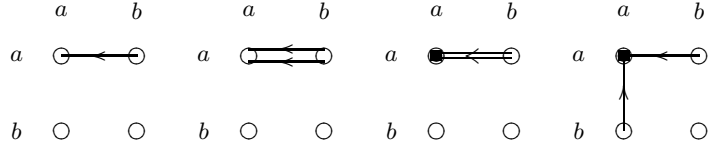

\begin{center}
\begin{tabular}{cccc}
\rxya{0.7}{
,(5,-5);(10,-5)**\dir{-}?(0.4)*\dir{<}
}
&
\;\;\;\;\;
\rxya{0.7}{
,(5,-4.6);(10,-4.6)**\dir{-}?(0.4)*\dir{<}
,(5,-5.4);(10,-5.4)**\dir{-}?(0.4)*\dir{<}
}
&
\;\;\;\;\;
\rxya{0.7}{
,(5,-5)*\cir(0.4,0){}*\frm{*}
,(5,-5);(10,-5)**\dir2{-}?(0.4)*\dir2{<}
}
&
\;\;\;\;\;
\rxya{0.7}{
,(5,-5)*\cir(0.4,0){}*\frm{*}
,(5,-5);(10,-5)**\dir{-}?(0.4)*\dir{<}
,(5,-5);(5,-10)**\dir{-}?(0.4)*\dir{<}
}
\\
\\
\end{tabular}
\end{center}
\caption{Four example diagrams}
\label{exkraj}
\end{figure}

According to these rules, we can omit the number
$\pm1,\pm2$ under the arrows like in figure \ref{exkraj}, since they are now redundant.
\vspace{1\baselineskip}

The easiest way to understand the reconstruction of a spectral triple from a diagram is by giving a few examples: \\  Take the algebra
$\mathcal{A}=\mathbb{H}\oplus M_3(\mathbb{C})\owns (a,b)$ with the first diagram of figure \ref{exkraj}.
Then the multiplicity matrix is
\bb
\mu=\pp{-1&1 \cr 0&0 }.
\eee
Using (\ref{representation}), its representation is, up to
unitary equivalence
\bb
\rho_L (a,b)=a \otimes 1_2,\, \rho _R (a,b) =b \otimes 1_2,\, \rho_L^c (a,b)=1_2\otimes
a,\,
\rho _R^c(a,b)=1_3\otimes a.
\eee The Hilbert space is
\bb
\mathcal{H}=\mathbb{C}^4 \oplus \mathbb{C}^6 \oplus \mathbb{C} ^4 \oplus \mathbb{C}^6.
\eee
In its Dirac operator (\ref{opdirac}), $\mathcal{M}=M\otimes 1_2$,
where $M$ is a nonvanishing complex $2\times 3 $
matrix.

\noindent Real structure, $S^0$-real structure and chirality are given by
(cc stands for
complex conjugation)
\bb J=
\begin{pmatrix} 0&1_{10}\\ 1_{10}&0
\end{pmatrix}
\circ {\rm cc},\quad
\epsilon =
\begin{pmatrix} 1_{10}&0\\ 0&-1_{10}
\end{pmatrix} ,\quad \chi =
\begin{pmatrix}-1_4&0&0&0\\ 0&1_6&0&0\\ 0&0&-1_4&0\\ 0&0&0&1_6
\end{pmatrix} .\eee The first tensor factor in $a\otimes 1_2$ concerns
particles, the second concerns antiparticles denoted by
$\cdot^c$.  The antiparticle representation is read from the transposed
multiplicity matrix.

For the second diagram of figure \ref{exkraj} we find the multiplicity matrix
\bb
\mu=\pp{-2&2 \cr 0&0 }.
\eee
The representation is read off as
\bb
\rho _L(a,b)=\pp{a&0\\ 0&a} \otimes 1_2,\quad \rho _R(a,b)=\pp{b&0\\ 0& b}\otimes 1_2,
\nonumber\\[2mm]
\rho _L^c(a,b)=\pp{1_2&0\\ 0&1_2}\otimes a,\quad
\rho _R^c(a,b)=\pp{1_3&0\\ 0&1_3}\otimes a.
\eee
The Hilbert space is
\bb
\mathcal{H}=\mathbb{C}^4 \oplus \mathbb{C}^4 \oplus \mathbb{C}^6 \oplus \mathbb{C}^6 \oplus \mathbb{C} ^4 \oplus \mathbb{C} ^4 \oplus \mathbb{C}^6 \oplus \mathbb{C}^6,
\eee
and for the mass matrix in the Dirac operator we find
\bb
\mathcal{M}=\pp{M_1&0\\ 0&M_2}\otimes 1_2,
\eee
where $M_1$ and $M_2$ are nonvanishing complex $2\times 3 $
matrices.
\noindent Real structure, $S^0$-real structure and chirality are given by
\bb J=
\begin{pmatrix} 0&1_{20}\\ 1_{20}&0
\end{pmatrix}
\circ {\rm cc},\quad
\epsilon =
\begin{pmatrix} 1_{20}&0\\ 0&-1_{20}
\end{pmatrix} ,\quad \chi =
\begin{pmatrix}-1_8&0&0&0\\ 0&1_{12}&0&0\\ 0&0&-1_8&0\\ 0&0&0&1_{12}
\end{pmatrix} .\eee
The third diagram of figure \ref{exkraj} yields
\bb
\mu=\pp{-1&2 \cr 0&0},
\eee
and its spectral triple reads:
\bb
\rho _L(a,b)=a\otimes 1_2,\quad \rho _R(a,b)=\pp{b&0\\ 0& b}\otimes 1_2,
\nonumber\\[2mm]
\rho _L^c(a,b)=1_2\otimes a,\quad
\rho _R^c(a,b)=\pp{1_3&0\\ 0&1_3}\otimes a, \\[2mm]
\mathcal{H}=\mathbb{C}^4  \oplus \mathbb{C}^6 \oplus \mathbb{C}^6 \oplus  \mathbb{C} ^4 \oplus \mathbb{C}^6 \oplus \mathbb{C}^6, \\[2mm]
\mathcal{M}=\pp{M_1&M_2}\otimes 1_2,\, M_1\ {\rm and}\ M_2
\ {\rm of \ size}\ 2\times 3,
\eee
\vspace{-0.9cm}
\bb
 J=\pp{0&1_{16}\\ 1_{16}&0}\circ {\rm cc},\quad
\epsilon =\pp{1_{16}&0\\ 0&-1_{16}},\quad \chi =\pp{-1_4&0&0&0\\
0&1_{12}&0&0\\ 0&0&-1_4&0\\ 0&0&0&1_{12}}.
\eee
Finally, still for the same algebra, let us consider the last diagram of
figure \ref{exkraj}. It gives
\bb
\mu=\pp{-1&1 \cr 1&0 },
\eee
and
\bb
\rho _L(a,b)=a\otimes 1_2,\quad \rho _R(a,b)=\pp{b\otimes 1_2&0\\ 0& a\otimes 1_3},
\nonumber \\[2mm]
\rho _L^c(a,b)=1_2\otimes a,\quad
\rho _R^c(a,b)=\pp{1_3\otimes a&0\\ 0&1_2\otimes b},\\[2mm]
\mathcal{H}=\mathbb{C}^4  \oplus \mathbb{C}^6 \oplus \mathbb{C}^6 \oplus  \mathbb{C} ^4 \oplus \mathbb{C}^6 \oplus \mathbb{C}^6, \\[2mm]
\mathcal{M}=\pp{M_1\otimes 1_2&1_2\otimes M_2},\quad M_1\ {\rm and}\ M_2
\ {\rm of \ size}\ 2\times 3.
\eee
In the four above examples all arrows have left algebra $\mathbb{H}$,
right algebra $M_3(\mathbb{C})$ and colour algebra
$\mathbb{H}$.

We can exhibit in these simple examples two features that will
become central for our algorithm. First, the distinction between a
double arrow in the third diagram and an ''edge'' in the fourth
diagram. Although transposing a single arrow amounts just to
exchanging particle and anti-particle and is thus of no physical
relevance, connecting arrows by building an edge alters the Dirac
operator, the Hilbert space and the representation fundamentally.
As we can immediately see from the mass matrices the spectral
triple with an edge cannot be converted into a spectral triple of
two connected parallel arrows. Only the simultaneous transposition
of all arrows connected at a point results in a mere exchange of
particles and antiparticles. Consequently we have to take
''edges'' like the fourth diagram into account separately.

Secondly we observe that the first diagram is minimal, if we delete an arrow the determinant of
the intersection form will become zero (in fact the whole intersection form will be zero).
We can erase an arrow from the other diagrams, dropping back to the first diagram (or its
transposed). These diagrams are therefore reducible to the first diagram.

On the other hand the second diagram is reducible to the third
diagram, since its spectral triples can be reduced by sizing down
the Hilbert space, the left representation and adjusting the mass
matrix. The last diagram has a similar relation to the second, but
in addition one of its arrows is transposed. It can thus not be
found by simply reducing the second diagram, and so we always have
to take the edges into account separately. We should perhaps note
here that in the case of three and more matrix algebras one arrow
will not be sufficient to produce a non-zero determinant of the
intersection form.

For the physical content of the diagrams, i.e. for the Yang-Mills-Higgs models they produce, it
is irrelevant if we reverse all the arrows at once, which is equivalent to multiplying the multiplicity
matrix with $-1$, or if we permute the algebras. Therefore diagrams
that differ only with respect to one of these operations will be considered equivalent. We will always
choose only one representative of the set of equivalent diagrams.

Arrows that are superfluous will not be taken into account either. These include two arrows connecting
the same algebras and having the same colour algebra but reversed directions. Their contributions
to the multiplicity matrix cancel out and thus the spectral triple is reducible. When several arrows
are connected, they have to be connected to the same point with common chirality. Otherwise three arrows would get connected
in a row and we could erase the middle arrow (i.e. the mass matrix) without altering the multiplicity
matrix.

We see that irreducible spectral triples are depicted by minimal (i.e. irreducible) Krajewski diagrams.
These have as few arrows
as possible which may be further reducible by connecting them or building edges. The aim of the algorithm
presented in this paper is to find all irreducible diagrams for a given number of matrix algebras. It should
be evident that the number of possibilities to put arrows in a diagram increases factorially with the number
of matrix algebras.

\section*{Appendix 4: Obtaining the Yang-Mills-Higgs theory}

To complete our short survey on the almost-commutative standard model, we will give a brief glimpse
on how to construct the actual Yang-Mills-Higgs theory.
We started out with the fixed (for convenience flat) Dirac operator of a 4-dimensional spacetime with a fixed fermionic
mass matrix. To generate curvature we have to perform a general coordinate transformation and then
fluctuate the Dirac operator. This can be achieved by lifting the automorphisms of the algebra to
the Hilbert space, unitarily transforming the Dirac operator with the lifted automorphisms
and then building linear combinations. Again we restrict ourselves to the finite case.
Except for complex conjugation in $M_n(\mathbb{C})$ and permutations of
identical summands in the algebra $\mathcal{A}=\mathcal{A}_1\oplus\mathcal{A}_2\oplus ...\oplus\mathcal{A}_N$,
every algebra automorphism
$\sigma
$  is inner, $\sigma (a)=uau^{-1}$ for a unitary $ u\in U(\mathcal{A})$. Therefore
the connected component of the automorphism group is
Aut$(\mathcal{A})^e=U(\mathcal{A})/(U(\mathcal{A})\cap{\rm Center}(\mathcal{A}))$. Its lift to the Hilbert
space \cite{real}
\bb L(\sigma )=\rho (u)J\rho (u)J^{-1}\eee is multi-valued.

The {\it fluctuation $\ddf$} of the Dirac operator $\mathcal{D}$ is given by a
finite collection $f$ of real numbers
$r_j$ and algebra automorphisms $\sigma _j\in{\rm Aut}(\mathcal{A})^e$ such
that
\bb
\ddf :=\sum_j r_j\,L(\sigma _j) \, \mathcal{D} \, L(\sigma_j)^{-1},\quad r_j\in\mathbb{R},\
\sigma _j\in{\rm Aut}(\mathcal{A})^e.
\eee
The fluctuated Dirac operator $\ddf$ is often denoted by $\varphi $, the
`Higgs scalar', in the physics literature.  We consider only fluctuations
with real coefficients since
$\ddf$ must remain selfadjoint.

To avoid the multi-valuedness in the fluctuations, we allow the entire
unitary group viewed as a (maximal) central extension of the
automorphism group.

As mentioned in the introduction an almost commutative geometry is the tensor product of a finite
noncommutative triple with an infinite, commutative spectral triple. By
Connes' reconstruction theorem \cite{grav} we know that the latter comes
from a Riemannian spin manifold, which we will take to be any
4-dimensional, compact, flat manifold like the flat 4-torus.  The spectral
action of this almost commutative spectral triple reduced to the finite part
is a functional on the vector space of all fluctuated, finite Dirac operators:
\bb V(\ddf )= \lambda\  \T\!\left[ (\ddf )^4\right] -\frac{\mu
^2}{2}\
\T\!\left[
(\ddf) ^2\right] ,\eee where $\lambda $ and $\mu $ are positive constants
\cite{cc}.
The spectral action is invariant under lifted automorphisms and even
under the unitary group $U(\mathcal{A})\owns u$,
\bb V( [\rho (u)J\rho (u)J^{-1}] \, \ddf \, [\rho (u)J\rho
(u)J^{-1}]^{-1})=V(\ddf),\eee and it is bounded from below.
To obtain the physical content of a diagram and its associated spectral triple one has
to find the minima $\ddfm $ of this action and their spectra. But this goes far beyond
the scope of this paper and we will content ourselves here with the algorithm to
find irreducible Krajewski diagrams.

\section*{Appendix 5: Deriving the spectral triple of AC-fermions}

The Krajewski diagram of the particle model under consideration encodes an almost-commutative spectral triple with six summands in the internal algebra:

\begin{center}
\begin{tabular}{c}
\rxyz{0.4}{
,(10,-10);(15,-10)**\dir{-}?(.6)*\dir{>}
,(5,-20);(15,-20)**\crv{(10,-17)}?(.6)*\dir{>}
,(10,-20);(15,-20)**\dir{-}?(.6)*\dir{>}
,(15,-20)*\cir(0.2,0){}*\frm{*}
,(30,-25);(25,-25)**\dir{-}?(.6)*\dir{>}
,(40,-35);(35,-35)**\dir{-}?(.6)*\dir{>}
} \\
\\
\end{tabular}
\end{center}
All the necessary translation rules between Krajewski
diagrams and the corresponding spectral triples can be found in \cite{Kraj} and \cite{1}.
The matrix is already {\it blown up} in the sense that the
representations of the complex parts of the matrix algebra have been fixed.
It is the same diagram from which the AC-fermions model of \cite{5} was derived.
One can clearly see the sub-diagram of the standard model in the upper $4\times 4$ corner.

To incorporate a new interaction for the AC-fermions the simplest possible extension
of the almost-commutative spectral triple \cite{5} will be to extend the quaternion
algebra $\mathbb{H}$ to the algebra of complex $2 \times 2$-matrices,
$M_2 (\mathbb{C})$. The notation for the algebra and its elements will be
the following:
\bb
\mathcal{A}= \mathbb{C} \oplus M_2(\mathbb{C}) \oplus M_3(\mathbb{C}) \oplus \mathbb{C} \oplus \mathbb{C} \oplus \mathbb{C} \ni (a,b,c,d,e,f),
\nonumber
\ee
which has as its representation
\bb
\rho_L (a,b,c,d,e,f)&=&  \pp{ b \otimes 1_3 & 0 & 0 & 0 \\ 0 & b & 0 & 0 \\ 0 & 0 & d & 0 \\ 0& 0& 0& \bar{e} },
\quad
\rho_R (a,b,c,d,e,f)=  \pp{ a 1_3 & 0 & 0 & 0 & 0 \\ 0 & \bar{a} 1_3 & 0 & 0 & 0\\ 0 & 0 & \bar{a}  & 0 & 0 \\ 0& 0& 0& e & 0 \\
0&0&0&0&f },
\nonumber \\ \nonumber
\ee

\bb
\rho_L^c (a,b,c,d,e,f)&=&  \pp{  1_2 \otimes c & 0 & 0 & 0 \\ 0 & \bar{a} 1_2 & 0 & 0 \\ 0 & 0 & d & 0 \\ 0& 0& 0& e },
\quad
\rho_R^c (a,b,c,d,e,f)=  \pp{ c & 0 & 0 & 0 & 0 \\ 0 & c & 0 & 0 & 0\\ 0 & 0 & \bar{a}  & 0 & 0 \\ 0& 0& 0& e & 0 \\
0&0&0&0& \bar{e}} .
\nonumber
\ee
These representations are faithful on the Hilbert space given below and serve as well to construct the lift of the automorphism group.
Roughly spoken each diagonal entry of the
representation of the algebra can be associated to fermion multiplet. For example the first entry of $\rho_L$, $b \otimes 1_3$,
is the representation of the algebra on the up and down quark doublet, where each quark is again a colour triplet.
As pointed out in \cite{farewell} the commutative sub-algebras of $\mathcal{A}$ which are equivalent to the complex numbers, serve
as receptacles for the $U(1)$ subgroups embedded in the automorphism group $U(2)\times U(3)$ of the $M_2(\mathbb(C) \oplus M_3(\mathbb{C})$ matrix algebra. One
can easily see that in contrast to the matrix algebra considered in \cite{5} there is now
a second $U(1)$ embedded in the automorphism group of the algebra. This new
$U(1)$ will be coupled to the AC-fermions only, as a minimal extension of the
standard model gauge group.
The extended lift is defined by
\bb
L(v,w) := \rho(\hat u , \hat v, \hat w, \hat x, \hat y, \hat z) J \rho(\hat u , \hat v, \hat w, \hat x, \hat y, \hat z) J^{-1},
\label{Lift}
\ee
with
\bb
\rho (a,b,c,d,e,f)  := \rho_L (a,b,c,d,e,f) \oplus \rho_R (a,b,c,d,e,f)
\oplus \rho_L^c (a,b,c,d,e,f) \oplus \rho_R^c (a,b,c,d,e,f),
\nonumber
\ee
where $J$ in (\ref{Lift}) is the real structure of the spectral triple, an anti-unitary
operator which coincides with the charge conjugation operator.
For the central extension the unitary entries in $L(u,v)$ are defined as
\bb
\hat u &:=&  (\det v )^{p_1} (\det w)^{q_1}  \nonumber \\
\hat v &:=&  v (\det v )^{p_2} (\det w)^{q_2}  \nonumber \\
\hat w &:=&  w (\det v )^{p_3} (\det w)^{q_3}   \nonumber \\
\hat x &:=&  (\det v )^{p_4} (\det w)^{q_4}   \nonumber \\
\hat y &:=&  (\det v )^{p_5} (\det w)^{q_5}  \nonumber \\
\hat z &:=&  (\det v )^{p_6} (\det w)^{q_6}  \nonumber
\ee
and the unitaries $(v,w) \in U(M_2 (\mathbb{C}) \oplus M_3(\mathbb{C}))$.
The exponents, or central charges, of the determinants will constitute the hypercharges corresponding
to the $U(1)$ subgroups of the gauge group. To ensure the absence of harmful
anomalies a rather cumbersome calculation, \cite{Anomaly,ZouZou}, results in the following values
for the central charges:
\bb
&p_1& \in \mathbb{Q}, \;\; q_1 \in \mathbb{Q},
\nonumber  \\
&p_2& = -\frac{1}{2}, \;\; q_2 = 0,
\nonumber  \\
&p_3& = \frac{p_1}{3}, \;\; q_3 = \frac{q_1 -1}{3},
\nonumber  \\
&p_4& \in \mathbb{Q}, \;\; q_4 \in \mathbb{Q},
\nonumber  \\
&p_5& = p_4, \; \; q_5 = q_4,
\nonumber  \\
&p_6& = - p_4, \;\; q_6 = - q_4
\nonumber
\ee
In the spirit of a minimal extension of the standard model with
AC-fermions as presented in \cite{leptons} the particle content of
the model should stay unchanged. Furthermore the standard model
fermions should not acquire any new interactions on tree-level. To
obtain the standard model hypercharge $U_Y(1)$ one can choose the
relevant central charges to be $p_1 := 0$ and $q_1 := - 1/2$.
Setting $q_4 := -1$ will produce electro-magnetic charges of $\mp
2$ for the AC-fermions $A^{--}$ and $C^{++}$, as required by
\cite{leptons}.

Now $p_4$ can still be  chosen freely to be any
rational number. If $p_4$ is taken to be different from zero, the AC-fermions
will be furnished with a new interaction generated by the second $U(1)$ sub-group,
henceforth called $U_{AC}(1)$,
in the group of unitaries of the algebra. Since the AC-fermions do not couple
to the Higgs scalar this new gauge group will not be
effected by the Higgs mechanism and stays thus unbroken.
In the model considered in this paper $p_4 := -1$ for simplicity and the
whole gauge group, before and after symmetry breaking is given by
\bb
U_Y(1)\times SU_w(2)\times SU_c(3)  \times U_{AC}(1) \longrightarrow
U_{em}(1)\times SU_c(3)  \times U_{AC}(1) \nonumber
\ee
The Hilbert space is  the direct sum of the standard model Hilbert
space, for details see \cite{schuck}, and the Hilbert space
containing the AC-fermions $A^{--}$ and $C^{++}$, see \cite{5}:
\bb \mathcal{H} = \mathcal{H}_{SM} \oplus \mathcal{H}_{AC},
\nonumber \ee where \bb \mathcal{H}_{AC} \ni \pp{\psi_{A L} \\
\psi_{C L}} \oplus \pp{\psi_{A R} \\ \psi_{C R}} \oplus
\pp{\psi_{A L}^c \\ \psi_{C L}^c} \oplus \pp{\psi_{A R}^c \\
\psi_{C R}^c}. \nonumber \ee
The wave functions $\psi_{A L}$, $\psi_{C L}$, $\psi_{A R}$ and
$\psi_{C R}$ are the respective left and right handed Dirac
4-spinors. The initial internal Dirac operator,  which is to be
fluctuated with the lifted automorphisms is chosen to be the mass
matrix
\bb \mathcal{M}= \pp{ \pp{m_u & 0 \\ 0 & m_d} \otimes 1_3 &0&0&0
&0\\ 0&0&m_e&0&0 \\ 0&0&0&m_A&0 \\ 0&0&0&0&m_C }, \nonumber \ee
with $m_u,m_d,m_e,m_A,m_C \in \mathbb{C}$.

It should be pointed out that the above choice of the summands of the matrix algebra,
the Hilbert space and the Dirac operator
is rather unique, if one
requires the Hilbert space to be minimal and the fermion masses to be non-degenerate.
Fluctuating the Dirac operator and calculating the spectral action gives the usual Einstein-Hilbert action,
the Yang-Mills-Higgs
action of the standard model and a new part in the Lagrangian for the two AC-fermions
as well as a term for the standard gauge potential $\tilde B_{\mu \nu}$ of the new
$U_{AC}(1)$ sub-group:
\bb \mathcal{L}_{AC} &=& i \psi_{A L}^\ast D_A \psi_{A L} + i
\psi_{A R}^\ast D_A \psi_{A R} + m_A \psi_{A L}^\ast \psi_{A R} +
 m_A \psi_{A R}^\ast \psi_{A L}
\nonumber  \\
&&+ \; i \psi_{C L}^\ast D_C \psi_{C L} + i \psi_{C R}^\ast D_C
\psi_{C R} + m_C \psi_{C L}^\ast \psi_{C R} + m_C \psi_{C R}^\ast
\psi_{C L}
\nonumber \\
&& \; - \frac{1}{4} \tilde B_{\mu \nu} \tilde B^{\mu \nu}.
\nonumber
\ee
The covariant derivative couples the AC-fermions to the $U(1)_Y$ sub-group of the standard model gauge group
and to the $U_{AC}(1)$ sub-group,
\bb D_{A/C} &=& \gamma^\mu \partial_\mu + \frac{i}{2} \, g' \,
Y_{A/C} \gamma^\mu B_\mu +  \frac{i}{2} \, g_{AC} \, \tilde
Y_{A/C}  \gamma^\mu \tilde B_\mu
\nonumber \\
&=&\gamma^\mu  \partial_\mu + \frac{i}{2} \, e \, Y_{A/C}
\gamma^\mu  A_\mu - \frac{i}{2} \, g' \, \sin \theta_w Y_{A/C}
\gamma^\mu  Z_\mu +  \frac{i}{2} \, g_{AC} \, \tilde Y_{A/C}
\gamma^\mu \tilde B_\mu, \nonumber \ee
where $\tilde B$ is the gauge field corresponding to $U(1)_{AC}$,
$g_{AC}$ is the corresponding coupling and $ \tilde Y_{A/C} $ the
almost-commutative hypercharge. From $\tilde Y_A = -\tilde Y_C= 2
p_4$ follows with the choice $p4=-1$ that $\tilde Y_A = - \tilde
Y_C = -2$. The possible range of the coupling $g_{AC}$ cannot be
given by almost-commutative geometry, but has to be fixed by
experiment. Furthermore $B$ is the gauge field corresponding to
$U(1)_Y$, $A$ and $Z$ are the photon and the Z-boson fields, $e$
is the electro-magnetic coupling and $\theta_w$ is the weak angle.
The hyper-charge $Y_{A/C}= 2 q_4$ of the AC-fermions can be any
non-zero fractional number with $Y_A = - Y_C$ so that $\psi_A$ and
$\psi_C$ have opposite electrical charge. To reproduce the
AC-model $q_4=-1$ was chosen, as stated above, and so  $Y_A=-2$
which results in opposite electro-magnetic charges $\mp 2 e$ for
the AC-fermions $A$ and $C$.

\bibliography{noncom,kosmo}

\begin{thebibliography}{48}
\expandafter\ifx\csname natexlab\endcsname\relax\def\natexlab#1{#1}\fi
\expandafter\ifx\csname bibnamefont\endcsname\relax
  \def\bibnamefont#1{#1}\fi
\expandafter\ifx\csname bibfnamefont\endcsname\relax
  \def\bibfnamefont#1{#1}\fi
\expandafter\ifx\csname citenamefont\endcsname\relax
  \def\citenamefont#1{#1}\fi
\expandafter\ifx\csname url\endcsname\relax
  \def\url#1{\texttt{#1}}\fi
\expandafter\ifx\csname urlprefix\endcsname\relax\def\urlprefix{URL }\fi
\providecommand{\bibinfo}[2]{#2}
\providecommand{\eprint}[2][]{\url{#2}}

\bibitem[{\citenamefont{Khlopov and Shibaev}(2002)}]{Shibaev}
\bibinfo{author}{\bibfnamefont{M.~Y.} \bibnamefont{Khlopov}} \bibnamefont{and}
  \bibinfo{author}{\bibfnamefont{K.~I.} \bibnamefont{Shibaev}},
  \bibinfo{journal}{Grav. Cosmol. Suppl.} \textbf{\bibinfo{volume}{8N1}},
  \bibinfo{pages}{45} (\bibinfo{year}{2002}).

\bibitem[{\citenamefont{Belotsky et~al.}(2000)\citenamefont{Belotsky, Khlopov,
  and Shibaev}}]{Sakhenhance}
\bibinfo{author}{\bibfnamefont{K.~M.} \bibnamefont{Belotsky}},
  \bibinfo{author}{\bibfnamefont{M.~Y.} \bibnamefont{Khlopov}},
  \bibnamefont{and} \bibinfo{author}{\bibfnamefont{K.~I.}
  \bibnamefont{Shibaev}}, \bibinfo{journal}{Grav. Cosmol. Suppl.}
  \textbf{\bibinfo{volume}{6}}, \bibinfo{pages}{140} (\bibinfo{year}{2000}).

\bibitem[{\citenamefont{Fargion et~al.}(1999)\citenamefont{Fargion, Golubkov,
  Khlopov, Konoplich, and Mignani}}]{Fargion99}
\bibinfo{author}{\bibfnamefont{D.}~\bibnamefont{Fargion}},
  \bibinfo{author}{\bibfnamefont{Y.~A.} \bibnamefont{Golubkov}},
  \bibinfo{author}{\bibfnamefont{M.~Y.} \bibnamefont{Khlopov}},
  \bibinfo{author}{\bibfnamefont{R.~V.} \bibnamefont{Konoplich}},
  \bibnamefont{and} \bibinfo{author}{\bibfnamefont{R.}~\bibnamefont{Mignani}},
  \bibinfo{journal}{Pisma Zh. Eksp. Teor. Fiz.} \textbf{\bibinfo{volume}{69}},
  \bibinfo{pages}{402} (\bibinfo{year}{1999}), \eprint{astro-ph/9903086}.

\bibitem[{\citenamefont{Fargion et~al.}(2000)\citenamefont{Fargion, Konoplich,
  Grossi, and Khlopov}}]{Grossi}
\bibinfo{author}{\bibfnamefont{D.}~\bibnamefont{Fargion}},
  \bibinfo{author}{\bibfnamefont{R.}~\bibnamefont{Konoplich}},
  \bibinfo{author}{\bibfnamefont{M.}~\bibnamefont{Grossi}}, \bibnamefont{and}
  \bibinfo{author}{\bibfnamefont{M.}~\bibnamefont{Khlopov}},
  \bibinfo{journal}{Astropart. Phys.} \textbf{\bibinfo{volume}{12}},
  \bibinfo{pages}{307} (\bibinfo{year}{2000}), \eprint{astro-ph/9809260}.

\bibitem[{\citenamefont{Belotsky and Khlopov}(2002)}]{Belotsky}
\bibinfo{author}{\bibfnamefont{K.~M.} \bibnamefont{Belotsky}} \bibnamefont{and}
  \bibinfo{author}{\bibfnamefont{M.~Y.} \bibnamefont{Khlopov}},
  \bibinfo{journal}{Grav. Cosmol. Suppl.} \textbf{\bibinfo{volume}{8N1}},
  \bibinfo{pages}{112} (\bibinfo{year}{2002}).

\bibitem[{\citenamefont{Belotsky and Khlopov}(2001)}]{BKS}
\bibinfo{author}{\bibfnamefont{K.~M.} \bibnamefont{Belotsky}} \bibnamefont{and}
  \bibinfo{author}{\bibfnamefont{M.~Y.} \bibnamefont{Khlopov}},
  \bibinfo{journal}{Grav. Cosmol.} \textbf{\bibinfo{volume}{7}},
  \bibinfo{pages}{183} (\bibinfo{year}{2001}).

\bibitem[{\citenamefont{Berezhiani and Nardi}(1995)}]{Berezhiani:1995du}
\bibinfo{author}{\bibfnamefont{Z.}~\bibnamefont{Berezhiani}} \bibnamefont{and}
  \bibinfo{author}{\bibfnamefont{E.}~\bibnamefont{Nardi}},
  \bibinfo{journal}{Phys. Lett.} \textbf{\bibinfo{volume}{B355}},
  \bibinfo{pages}{199} (\bibinfo{year}{1995}), \eprint{hep-ph/9503367}.

\bibitem[{\citenamefont{Maltoni et~al.}(2000)\citenamefont{Maltoni, Novikov,
  Okun, Rozanov, and Vysotsky}}]{Okun}
\bibinfo{author}{\bibfnamefont{M.}~\bibnamefont{Maltoni}},
  \bibinfo{author}{\bibfnamefont{V.~A.} \bibnamefont{Novikov}},
  \bibinfo{author}{\bibfnamefont{L.~B.} \bibnamefont{Okun}},
  \bibinfo{author}{\bibfnamefont{A.~N.} \bibnamefont{Rozanov}},
  \bibnamefont{and} \bibinfo{author}{\bibfnamefont{M.~I.}
  \bibnamefont{Vysotsky}}, \bibinfo{journal}{Phys. Lett.}
  \textbf{\bibinfo{volume}{B476}}, \bibinfo{pages}{107} (\bibinfo{year}{2000}),
  \eprint{hep-ph/9911535}.

\bibitem[{\citenamefont{Volovik}(2003)}]{Volovik:2003kh}
\bibinfo{author}{\bibfnamefont{G.~E.} \bibnamefont{Volovik}},
  \bibinfo{journal}{Pisma Zh. Eksp. Teor. Fiz.} \textbf{\bibinfo{volume}{78}},
  \bibinfo{pages}{1203} (\bibinfo{year}{2003}), \eprint{hep-ph/0310006}.

\bibitem[{\citenamefont{Belotsky et~al.}(2004)}]{4had}
\bibinfo{author}{\bibfnamefont{K.}~\bibnamefont{Belotsky}} \bibnamefont{et~al.}
  (\bibinfo{year}{2004}), \eprint{hep-ph/0411271}.

\bibitem[{\citenamefont{Khlopov}(2006)}]{anutium}
\bibinfo{author}{\bibfnamefont{M.~Y.} \bibnamefont{Khlopov}},
  \bibinfo{journal}{JETP Lett.} \textbf{\bibinfo{volume}{83}},
  \bibinfo{pages}{1} (\bibinfo{year}{2006}), \eprint{astro-ph/0511796}.

\bibitem[{\citenamefont{Glashow}(2005)}]{Glashow}
\bibinfo{author}{\bibfnamefont{S.~L.} \bibnamefont{Glashow}}
  (\bibinfo{year}{2005}), \eprint{hep-ph/0504287}.

\bibitem[{\citenamefont{Fargion and Khlopov}(2005)}]{Fargion:2005xz}
\bibinfo{author}{\bibfnamefont{D.}~\bibnamefont{Fargion}} \bibnamefont{and}
  \bibinfo{author}{\bibfnamefont{M.}~\bibnamefont{Khlopov}}
  (\bibinfo{year}{2005}), \eprint{hep-ph/0507087}.

\bibitem[{\citenamefont{Fargion et~al.}(2005)\citenamefont{Fargion, Khlopov,
  and Stephan}}]{leptons}
\bibinfo{author}{\bibfnamefont{D.}~\bibnamefont{Fargion}},
  \bibinfo{author}{\bibfnamefont{M.}~\bibnamefont{Khlopov}}, \bibnamefont{and}
  \bibinfo{author}{\bibfnamefont{C.~A.} \bibnamefont{Stephan}}
  (\bibinfo{year}{2005}), \eprint{astro-ph/0511789}.

\bibitem[{\citenamefont{Stephan}(2005)}]{5}
\bibinfo{author}{\bibfnamefont{C.~A.} \bibnamefont{Stephan}}
  (\bibinfo{year}{2005}), \eprint{hep-th/0509213}.

\bibitem[{\citenamefont{Connes}(1994)}]{book}
\bibinfo{author}{\bibfnamefont{A.}~\bibnamefont{Connes}},
  \emph{\bibinfo{title}{Noncommutative geometry}} (\bibinfo{publisher}{Academic
  Press}, \bibinfo{address}{London and San Diego}, \bibinfo{year}{1994}).

\bibitem[{\citenamefont{Iochum et~al.}(2004)\citenamefont{Iochum, Sch{\"u}cker,
  and Stephan}}]{1}
\bibinfo{author}{\bibfnamefont{B.}~\bibnamefont{Iochum}},
  \bibinfo{author}{\bibfnamefont{T.}~\bibnamefont{Sch{\"u}cker}},
  \bibnamefont{and} \bibinfo{author}{\bibfnamefont{C.~A.}
  \bibnamefont{Stephan}}, \bibinfo{journal}{J. Math. Phys.}
  \textbf{\bibinfo{volume}{45}}, \bibinfo{pages}{5003} (\bibinfo{year}{2004}),
  \eprint{hep-th/0312276}.

\bibitem[{\citenamefont{Gracia-Bondia et~al.}(2001)\citenamefont{Gracia-Bondia,
  Varilly, and Figueroa}}]{costa}
\bibinfo{author}{\bibfnamefont{J.~M.} \bibnamefont{Gracia-Bondia}},
  \bibinfo{author}{\bibfnamefont{J.~C.} \bibnamefont{Varilly}},
  \bibnamefont{and} \bibinfo{author}{\bibfnamefont{H.}~\bibnamefont{Figueroa}},
  \emph{\bibinfo{title}{Elements of noncommutative geometry}}
  (\bibinfo{publisher}{Birkhaeuser}, \bibinfo{address}{Boston},
  \bibinfo{year}{2001}).

\bibitem[{\citenamefont{Chamseddine and Connes}(1997)}]{cc}
\bibinfo{author}{\bibfnamefont{A.~H.} \bibnamefont{Chamseddine}}
  \bibnamefont{and} \bibinfo{author}{\bibfnamefont{A.}~\bibnamefont{Connes}},
  \bibinfo{journal}{Commun. Math. Phys.} \textbf{\bibinfo{volume}{186}},
  \bibinfo{pages}{731} (\bibinfo{year}{1997}), \eprint{hep-th/9606001}.

\bibitem[{\citenamefont{Sch{\"u}cker}(2005{\natexlab{a}})}]{schuck}
\bibinfo{author}{\bibfnamefont{T.}~\bibnamefont{Sch{\"u}cker}},
  \bibinfo{journal}{Lect. Notes Phys.} \textbf{\bibinfo{volume}{659}},
  \bibinfo{pages}{285} (\bibinfo{year}{2005}{\natexlab{a}}),
  \eprint{hep-th/0111236}.

\bibitem[{\citenamefont{Riemann}(1968)}]{Riemann}
\bibinfo{author}{\bibfnamefont{B.}~\bibnamefont{Riemann}},
  \bibinfo{journal}{Abhandlungen der K{\"o}niglichen Gesellschaft der
  Wissenschaften zu G{\"o}ttingen} \textbf{\bibinfo{volume}{13}}
  (\bibinfo{year}{1968}).

\bibitem[{\citenamefont{Gelfand and Naimark}(1943)}]{Gelfand}
\bibinfo{author}{\bibfnamefont{I.}~\bibnamefont{Gelfand}} \bibnamefont{and}
  \bibinfo{author}{\bibfnamefont{M.}~\bibnamefont{Naimark}},
  \bibinfo{journal}{Mat. Sbornik} \textbf{\bibinfo{volume}{12}},
  \bibinfo{pages}{197} (\bibinfo{year}{1943}).

\bibitem[{\citenamefont{O'Raifeartaigh}(1965)}]{ORaif}
\bibinfo{author}{\bibfnamefont{L.}~\bibnamefont{O'Raifeartaigh}},
  \bibinfo{journal}{Phys. Rev.} \textbf{\bibinfo{volume}{139}},
  \bibinfo{pages}{1052} (\bibinfo{year}{1965}).

\bibitem[{\citenamefont{Coleman and Mandula}(1967)}]{Coleman}
\bibinfo{author}{\bibfnamefont{S.}~\bibnamefont{Coleman}} \bibnamefont{and}
  \bibinfo{author}{\bibfnamefont{J.}~\bibnamefont{Mandula}},
  \bibinfo{journal}{Phys. Rev.} \textbf{\bibinfo{volume}{159}},
  \bibinfo{pages}{1251} (\bibinfo{year}{1967}).

\bibitem[{\citenamefont{Mather}(1974)}]{Mather}
\bibinfo{author}{\bibfnamefont{J.}~\bibnamefont{Mather}},
  \bibinfo{journal}{Bull. Amerik. Math. Soc.} \textbf{\bibinfo{volume}{80}},
  \bibinfo{pages}{271} (\bibinfo{year}{1974}).

\bibitem[{\citenamefont{Paschke and Sitarz}(1998)}]{pasch}
\bibinfo{author}{\bibfnamefont{M.}~\bibnamefont{Paschke}} \bibnamefont{and}
  \bibinfo{author}{\bibfnamefont{A.}~\bibnamefont{Sitarz}},
  \bibinfo{journal}{J. Math. Phys.} \textbf{\bibinfo{volume}{39}},
  \bibinfo{pages}{6191} (\bibinfo{year}{1998}).

\bibitem[{\citenamefont{Krajewski}(1998)}]{Kraj}
\bibinfo{author}{\bibfnamefont{T.}~\bibnamefont{Krajewski}},
  \bibinfo{journal}{J. Geom. Phys.} \textbf{\bibinfo{volume}{28}},
  \bibinfo{pages}{1} (\bibinfo{year}{1998}), \eprint{hep-th/9701081}.

\bibitem[{\citenamefont{Jureit and Stephan}(2005)}]{2}
\bibinfo{author}{\bibfnamefont{J.-H.} \bibnamefont{Jureit}} \bibnamefont{and}
  \bibinfo{author}{\bibfnamefont{C.~A.} \bibnamefont{Stephan}},
  \bibinfo{journal}{J. Math. Phys.} \textbf{\bibinfo{volume}{46}},
  \bibinfo{pages}{043512} (\bibinfo{year}{2005}), \eprint{hep-th/0501134}.

\bibitem[{\citenamefont{Sch{\"u}cker}(2005{\natexlab{b}})}]{3}
\bibinfo{author}{\bibfnamefont{T.}~\bibnamefont{Sch{\"u}cker}}
  (\bibinfo{year}{2005}{\natexlab{b}}), \eprint{hep-th/0501181}.

\bibitem[{\citenamefont{Jureit et~al.}(2005)\citenamefont{Jureit, Sch{\"u}cker,
  and Stephan}}]{4}
\bibinfo{author}{\bibfnamefont{J.-H.} \bibnamefont{Jureit}},
  \bibinfo{author}{\bibfnamefont{T.}~\bibnamefont{Sch{\"u}cker}},
  \bibnamefont{and} \bibinfo{author}{\bibfnamefont{C.~A.}
  \bibnamefont{Stephan}}, \bibinfo{journal}{J. Math. Phys.}
  \textbf{\bibinfo{volume}{46}}, \bibinfo{pages}{072303}
  (\bibinfo{year}{2005}), \eprint{hep-th/0503190}.

\bibitem[{\citenamefont{Lazzarini and Sch{\"u}cker}(2001)}]{farewell}
\bibinfo{author}{\bibfnamefont{S.}~\bibnamefont{Lazzarini}} \bibnamefont{and}
  \bibinfo{author}{\bibfnamefont{T.}~\bibnamefont{Sch{\"u}cker}},
  \bibinfo{journal}{Phys. Lett.} \textbf{\bibinfo{volume}{B510}},
  \bibinfo{pages}{277} (\bibinfo{year}{2001}), \eprint{hep-th/0104038}.

\bibitem[{\citenamefont{Zeldovich and Khlopov}(1978)}]{ZK}
\bibinfo{author}{\bibfnamefont{Y.~B.} \bibnamefont{Zeldovich}}
  \bibnamefont{and} \bibinfo{author}{\bibfnamefont{M.~Y.}
  \bibnamefont{Khlopov}}, \bibinfo{journal}{Phys. Lett.}
  \textbf{\bibinfo{volume}{B79}}, \bibinfo{pages}{239} (\bibinfo{year}{1978}).

\bibitem[{\citenamefont{Dubrovich et~al.}(2004)\citenamefont{Dubrovich,
  Fargion, and Khlopov}}]{DFK}
\bibinfo{author}{\bibfnamefont{V.~K.} \bibnamefont{Dubrovich}},
  \bibinfo{author}{\bibfnamefont{D.}~\bibnamefont{Fargion}}, \bibnamefont{and}
  \bibinfo{author}{\bibfnamefont{M.~Y.} \bibnamefont{Khlopov}},
  \bibinfo{journal}{Astropart. Phys.} \textbf{\bibinfo{volume}{22}},
  \bibinfo{pages}{183} (\bibinfo{year}{2004}), \eprint{hep-ph/0312105}.

\bibitem[{\citenamefont{Wandelt et~al.}(2000)}]{McGuire:2001qj}
\bibinfo{author}{\bibfnamefont{B.~D.} \bibnamefont{Wandelt}}
  \bibnamefont{et~al.} (\bibinfo{year}{2000}), \eprint{astro-ph/0006344}.

\bibitem[{\citenamefont{Dover et~al.}(1979)\citenamefont{Dover, Gaisser, and
  Steigman}}]{Starkman}
\bibinfo{author}{\bibfnamefont{C.~B.} \bibnamefont{Dover}},
  \bibinfo{author}{\bibfnamefont{T.~K.} \bibnamefont{Gaisser}},
  \bibnamefont{and} \bibinfo{author}{\bibfnamefont{G.}~\bibnamefont{Steigman}},
  \bibinfo{journal}{Phys. Rev. Lett.} \textbf{\bibinfo{volume}{42}},
  \bibinfo{pages}{1117} (\bibinfo{year}{1979}).

\bibitem[{\citenamefont{McCammon et~al.}(2002)}]{XQC}
\bibinfo{author}{\bibfnamefont{D.}~\bibnamefont{McCammon}}
  \bibnamefont{et~al.}, \bibinfo{journal}{Astrophys. J.}
  \textbf{\bibinfo{volume}{576}}, \bibinfo{pages}{188} (\bibinfo{year}{2002}),
  \eprint{astro-ph/0205012}.

\bibitem[{\citenamefont{Khlopov}(1981)}]{fractons}
\bibinfo{author}{\bibfnamefont{M.~Y.} \bibnamefont{Khlopov}},
  \bibinfo{journal}{JETP Lett.} \textbf{\bibinfo{volume}{33}},
  \bibinfo{pages}{162} (\bibinfo{year}{1981}).

\bibitem[{\citenamefont{Fargion}(2004)}]{FarSolar03}
\bibinfo{author}{\bibfnamefont{D.}~\bibnamefont{Fargion}},
  \bibinfo{journal}{JHEP} \textbf{\bibinfo{volume}{06}}, \bibinfo{pages}{045}
  (\bibinfo{year}{2004}), \eprint{hep-ph/0312011}.

\bibitem[{\citenamefont{Fargion}(2002)}]{FarTau02}
\bibinfo{author}{\bibfnamefont{D.}~\bibnamefont{Fargion}},
  \bibinfo{journal}{Astrophys. J.} \textbf{\bibinfo{volume}{570}},
  \bibinfo{pages}{909} (\bibinfo{year}{2002}), \eprint{astro-ph/0002453}.

\bibitem[{\citenamefont{Fargion et~al.}(2004)\citenamefont{Fargion,
  De~Sanctis~Lucentini, and De~Santis}}]{FarTau03}
\bibinfo{author}{\bibfnamefont{D.}~\bibnamefont{Fargion}},
  \bibinfo{author}{\bibfnamefont{P.~G.} \bibnamefont{De~Sanctis~Lucentini}},
  \bibnamefont{and}
  \bibinfo{author}{\bibfnamefont{M.}~\bibnamefont{De~Santis}},
  \bibinfo{journal}{Astrophys. J.} \textbf{\bibinfo{volume}{613}},
  \bibinfo{pages}{1285} (\bibinfo{year}{2004}), \eprint{hep-ph/0305128}.

\bibitem[{\citenamefont{Winkelmann et~al.}(2005)\citenamefont{Winkelmann,
  Bunkov, and Godfrin}}]{Winkelmann:2005un}
\bibinfo{author}{\bibfnamefont{C.~B.} \bibnamefont{Winkelmann}},
  \bibinfo{author}{\bibfnamefont{Y.~M.} \bibnamefont{Bunkov}},
  \bibnamefont{and} \bibinfo{author}{\bibfnamefont{H.}~\bibnamefont{Godfrin}},
  \bibinfo{journal}{Grav. Cosmol.} \textbf{\bibinfo{volume}{11}},
  \bibinfo{pages}{87} (\bibinfo{year}{2005}).

\bibitem[{\citenamefont{Fargion et~al.}(2006)\citenamefont{Fargion, Khlopov,
  and Stephan}}]{CQG}
\bibinfo{author}{\bibfnamefont{D.}~\bibnamefont{Fargion}},
  \bibinfo{author}{\bibfnamefont{M.~Y.} \bibnamefont{Khlopov}},
  \bibnamefont{and} \bibinfo{author}{\bibfnamefont{C.~A.}
  \bibnamefont{Stephan}}, \bibinfo{journal}{Class. Quantum Grav.}
  \textbf{\bibinfo{volume}{23}}, \bibinfo{pages}{7305} (\bibinfo{year}{2006}).

\bibitem[{\citenamefont{Connes}(1995)}]{real}
\bibinfo{author}{\bibfnamefont{A.}~\bibnamefont{Connes}}, \bibinfo{journal}{J.
  Math. Phys.} \textbf{\bibinfo{volume}{36}}, \bibinfo{pages}{6194}
  (\bibinfo{year}{1995}).

\bibitem[{\citenamefont{Connes}(1996)}]{grav}
\bibinfo{author}{\bibfnamefont{A.}~\bibnamefont{Connes}},
  \bibinfo{journal}{Commun. Math. Phys.} \textbf{\bibinfo{volume}{182}},
  \bibinfo{pages}{155} (\bibinfo{year}{1996}), \eprint{hep-th/9603053}.

\bibitem[{\citenamefont{E.~Alvarez et~al.}(1995)\citenamefont{E.~Alvarez,
  Gracia-Bondia, and Martin}}]{Anomaly}
\bibinfo{author}{\bibfnamefont{J.}~\bibnamefont{E.~Alvarez}},
  \bibinfo{author}{\bibnamefont{Gracia-Bondia}}, \bibnamefont{and}
  \bibinfo{author}{\bibfnamefont{C.}~\bibnamefont{Martin}},
  \bibinfo{journal}{Phys.Lett.} \textbf{\bibinfo{volume}{B364}},
  \bibinfo{pages}{33} (\bibinfo{year}{1995}), \eprint{hep-th/9506115}.

\bibitem[{\citenamefont{Sch{\"u}cker and Zouzou}(2001)}]{ZouZou}
\bibinfo{author}{\bibfnamefont{T.}~\bibnamefont{Sch{\"u}cker}}
  \bibnamefont{and} \bibinfo{author}{\bibfnamefont{S.}~\bibnamefont{Zouzou}}
  (\bibinfo{year}{2001}), \eprint{hep-th/0109124}.

\bibitem[{\citenamefont{Jackiw}(2002)}]{Jackiw}
\bibinfo{author}{\bibfnamefont{R.}~\bibnamefont{Jackiw}},
  \bibinfo{journal}{Nucl. Phys. Proc. Suppl.} \textbf{\bibinfo{volume}{108}},
  \bibinfo{pages}{30} (\bibinfo{year}{2002}), \eprint{hep-th/0110057}.

\bibitem[{\citenamefont{Snyder}(1947)}]{Snyder}
\bibinfo{author}{\bibfnamefont{H.}~\bibnamefont{Snyder}},
  \bibinfo{journal}{Phys. Rev.} \textbf{\bibinfo{volume}{71}},
  \bibinfo{pages}{38} (\bibinfo{year}{1947}).

\end{thebibliography}

\end{document}